\documentclass[%
 reprint,
superscriptaddress,
 amsmath,amssymb,
 aps,
 prb,
]{revtex4-1}

\usepackage{graphicx}
\usepackage{dcolumn}
\usepackage{bm}
\usepackage{hyperref}

\newcommand{\calH}{\mathcal{H}}
\newcommand{\calO}{\mathcal{O}}
\newcommand{\calL}{\mathcal{L}}
\newcommand{\calJ}{\mathcal{J}}
\newcommand{\calD}{\mathcal{D}}

\newcommand{\eff}{\mathrm{eff}}
\usepackage{braket}
\usepackage{floatrow}
\usepackage[caption=false]{subfig}
\usepackage{here}

\begin{document}

\preprint{APS/123-QED}

\title{Laser-irradiated Kondo insulators : \\ Controlling Kondo effect and topological phases  }

\author{Kazuaki Takasan}
\email{takasan@scphys.kyoto-u.ac.jp}
\affiliation{%
 Department of Physics, Kyoto University, Kyoto 606-8502, Japan
}%
\author{Masaya Nakagawa}
\affiliation{%
 RIKEN Center for Emergent Matter Science (CEMS), Wako, Saitama 351-0198, Japan
}%
\author{Norio Kawakami}
\affiliation{%
 Department of Physics, Kyoto University, Kyoto 606-8502, Japan
}%

\date{\today}

\begin{abstract}
We theoretically investigate the nature of laser-irradiated Kondo insulators. Using Floquet theory and slave boson approach, we study a periodic Anderson model and derive an effective model which describes the laser-irradiated Kondo insulators. In this model, we find two generic effects induced by laser light.
One is the dynamical localization, which suppresses hopping and hybridization. 
The other is the laser-induced hopping and hybridization, which can be interpreted as a synthetic spin-orbit coupling or magnetic field.
The first effect drastically changes the behavior of the Kondo effect.
Especially, the Kondo effect under laser light qualitatively changes its character depending on whether the hybridization is on-site or off-site.
The second effect triggers topological phase transitions. 
In topological Kondo insulators, linearly polarized laser light realizes phase transitions between trivial, weak topological, and strong topological Kondo insulators.
Moreover, circularly polarized laser light breaks time-reversal symmetry and induces Weyl semimetallic phases.
Our results pave the new way to dynamically control the Kondo effect and topological phases in heavy fermion systems. 
We also discuss experimental setups to detect the signatures.
\end{abstract}

\maketitle
\section{Introduction}
Controlling the quantum states via the periodic driving is gathering great attentions in recent years.
It is called \textit{Floquet engineering}, which has been developed mainly in cold atomic systems \cite{EckardtRMP2017}.
In cold atomic physics, Floquet engineering has already been an important and established tool to realize the desired systems.
For example, topologically non-trivial states in the Haldane model\cite{Haldane1988} were experimentally realized in periodically driven honeycomb lattice and have been providing a typical platform of investigating the physics of topological phases \cite{Jotzu2014, Flaschner2016},

As for solid states, important and interesting scenarios of Floquet engineering have been proposed in graphene and semiconductors \cite{Oka2009, Kitagawa2011, Lindner2011, Dora2012, Delplace2013, Ezawa2013, Rudner2013, Titum2016}.
For most of solid state systems, the time-periodic perturbations are due to the irradiation of strong laser light coupled to electrons or phonons. For example, it has been theoretically proposed that topologically non-trivial states of matter are realized in laser-irradiated semiconductor quantum wells \cite{Lindner2011}. Experimentally it has been confirmed that the band structure is indeed modified and Floquet states are realized on the surface states of a laser-irradiated topological insulator Bi$_2$Se$_3$ \cite{Wang2013, Mahmood2016}.
 Therefore, for weakly interacting systems, laser light is becoming a useful tool to control the quantum states of matter.
On the other hand, the investigation in strongly correlated electron systems is still limited \cite{Grushin2014, Mikami2016}. 
However, since various quantum phases of matter, such as magnetically ordered states or Mott insulating states, are realized only in strongly correlated systems, it is important to search for the possibility of Floquet engineering in strongly correlated electron systems.

\begin{figure}[tbp]
\includegraphics[width=5.5cm]{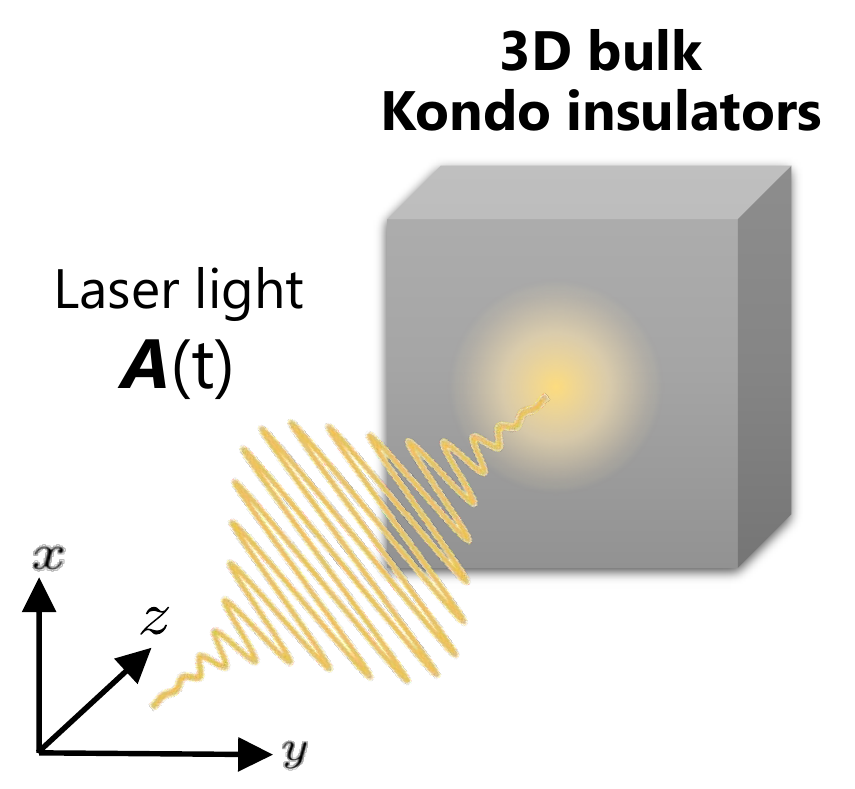}
\caption{Schematic picture of our setup. A three-dimensional bulk Kondo insulator is irradiated by laser light in the $z$-direction. We assume a pulse laser wide enough to apply Floquet theory which is a theoretical prescription for time-periodic systems. }
\label{setup}
\end{figure}

In this paper, we focus on the Floquet engineering in heavy fermion systems, which are a typical example of the strongly correlated electron systems\cite{Coleman_book}. Here the interplay between conduction electrons and localized electrons induces the Kondo effect and gives us rich phase diagrams including topological phases\cite{Dzero2010, Dzero2015}.
To be specific, we theoretically investigate the effect of  the laser irradiation on the periodic Anderson model. The schematic picture of the setup is shown in Fig. \ref{setup}.
The periodic Anderson model is a simple model of Kondo insulators, i.e. strongly correlated insulators induced by the Kondo effect \cite{Riseborough2000}.
Using Floquet theory and slave boson approach,
we derive a general effective model which describes the laser-irradiated Kondo insulators.
We show that there are two laser-induced effects in this model. One is the dynamical localization \cite{Dunlap1986}, which suppresses hopping and hybridization. The other is the laser-induced hopping and hybridization. These effects change the properties of the original Kondo insulators. The first effect, dynamical localization, modifies the behavior of the Kondo effect. 
Especially, we point out that the Kondo effect can be enhanced or suppressed by the laser light depending on the spatial structures of hybridization. 
The second effect, laser-induced hopping and hybridization, drives topological phase transitions.
In three dimensional topological Kondo insulators \cite{Dzero2010, Dzero2015}, linearly polarized laser light induces a phase transition between trivial, weak topological, and strong topological Kondo insulators. Furthermore, with circularly polarized laser light, the system realizes strongly correlated Weyl semimetals \cite{Wan2011, Murakami2007, Yan2017}. Our results pave the way to engineer the electronic properties dynamically in heavy fermion systems.

This paper is organized as follows.
In Sec. II, we introduce the periodic Anderson model and two theoretical methods. 
The first method is slave boson technique to treat the interaction effect. 
The second method is Floquet theory which is known as a versatile tool for time-periodic quantum systems. 
In Sec. III, we discuss our main results. We derive a general effective model that describes laser-irradiated Kondo insulators and show that there are two effects induced by laser light. 
To demonstrate how these effects change the nature of the original systems, we discuss the impact of them on the Kondo effect in Sec. IV. We then discuss how the laser-induced effects change the topological properties of the original systems in Sec.V. We show that these effects lead to topological phase transitions. Furthermore, we discuss experimental setups to confirm our results in Sec.VI. Finally summary and outlook are presented in Sec.VII.

\section{Model and Methods}

\subsection{Periodic Anderson model with slave boson approach}

In order to study the effect of laser light on Kondo insulators, 
we introduce a theoretical model, which is a variant of the periodic Anderson model.
The model Hamiltonian reads
\begin{align}
\calH&=\sum_{i j \sigma} \{ t_{c, i j}-\mu \delta_{ij} \}  c^\dagger_{i \sigma} c_{j \sigma} + \sum_{i j \sigma} 
\{t_{f, i j} + (\epsilon_f - \mu )\delta_{ij}   \} f^\dagger_{i \sigma} f_{j \sigma} \nonumber \\
&+\sum_{i j \sigma \sigma^\prime}\{ V_{i j \sigma \sigma^\prime} c^\dagger_{i \sigma} f_{j \sigma^\prime} + \mathrm{h.c.} \} +U \sum_i n_{i \uparrow}^{(f)} n_{i \downarrow}^{(f)},
\label{Model_PAM}
\end{align}
with $ t_{c, i i} = t_{f, i i} = 0$, $t_{c, i j}= (t_{c, j i})^*$ and $t_{f, i j}= (t_{f, j i})^*$. We assume $t_{f, ij} \ll t_{c, ij}$. $\sigma, \sigma^\prime (= \uparrow, \downarrow)$ stand for the (pseudo)spin. This model consists of conduction electrons specified by the annihilation (creation) operator $c_{i \sigma}$ ($c^\dagger_{i \sigma}$) and almost localized $f$-electrons specified by $f_{i \sigma}$ ($f^\dagger_{i \sigma}$). 
The term including $V_{i j \sigma \sigma^\prime}$ represents the hybridization of the conduction and localized $f$-electrons.
Due to the localized orbit of $f$-electrons, the electron correlation among $f$-electrons is strong, and thus we introduce a Hubbard-type interaction for them, as shown in the last term of the Hamiltonian (\ref{Model_PAM}). $n_{i \sigma}^{(f)}$ represents the number operator of the $f$-orbit at the $i$-th site and the definition is $n_{i \sigma}^{(f)}= f^\dagger_{i \sigma} f_{i \sigma}$.

To incorporate the correlation effect among $f$-electrons, 
we use a slave-boson mean field treatment \cite{Coleman1983, Coleman1987, Newns1987, Millis1987}.
It assumes a renormalized band structure, 
which results in the mean-field Hamiltonian,
\begin{align}
\calH_\mathrm{MF}&
=\sum_{i j \sigma} \{ t_{c, i j}-\mu \delta_{ij} \}  c^\dagger_{i \sigma} c_{j \sigma}\nonumber  \\
&+ \sum_{i j \sigma} \{|b|^2  t_{f, i j} + (\epsilon_f + \lambda - \mu )\delta_{ij}   \} f^\dagger_{i \sigma} f_{j \sigma} \nonumber \\
&+\sum_{i j \sigma \sigma^\prime}\{  b^*V_{i j \sigma \sigma^\prime} c^\dagger_{i \sigma} f_{j \sigma^\prime} + \mathrm{h.c.} \} ,
\label{Model_MF}
\end{align}
where $b$ is a renormalization factor and $\lambda$ represents an energy shift of {\it f}-orbital level by the interaction effect. 

In this study, we numerically solve self-consistent equations for $b$ and $\lambda$ and determine their values for each temperature. We show the derivation of the self-consistent equations in Appendix. A. The explicit form of the equations is 
\begin{align}
\lambda &= \frac{1}{N} \sum_{i \neq j \sigma} t_{f, i j} \langle f^\dagger_{i \sigma} f_{j \sigma} \rangle \nonumber \\ &\quad+ \frac{1}{N b^*} \sum_{i j \sigma \sigma^\prime} V^*_{ j i \sigma^\prime \sigma} \langle f^\dagger_{i \sigma} c_{j \sigma^\prime} \rangle, \label{SCeq1_0} \\
 |b|^2 &= 1 - \frac{1}{N} \sum_{i \sigma}\langle f^\dagger_{i \sigma} f_{i \sigma} \rangle \label{SCeq2_0},
\end{align}
where $N$ is the number of sites and $\langle \cdots \rangle$ stands for a thermal average with the mean-field Lagrangian $\calL_\mathrm{MF}$ (The definition is presented in Appendix. A). Since we consider insulating systems in this study, we solve simultaneously the equations (\ref{SCeq1_0}), (\ref{SCeq2_0}) and the condition of half-filling,
\begin{align}
n_\mathrm{fill}&=\frac{1}{N} \sum_{i \sigma} \left ( \langle c^\dagger_{i \sigma} c_{i \sigma} \rangle + \langle f^\dagger_{i \sigma} f_{i \sigma} \rangle \right )=2,
\end{align}
and determine the chemical potential $\mu$ for each temperature. 
Since there are two types of electrons ($c$ and $f$) and two (pseudo) spins ($\uparrow$ and $\downarrow$) for each sites, $n_\mathrm{fill} = 2$ corresponds to half-filling.

\subsection{Floquet theory}

Next we consider the effect of laser light. 
For this purpose, we first consider a time-dependent model which describes laser-illuminated Kondo insulators. 
We treat laser light as oscillating electric fields, and thus introduce a Peierls phase. 
We obtain the time-dependent model from (\ref{Model_MF}) as
\begin{align}
\calH (t)&
=\sum_{i j \sigma} \{e^{i \bm A (t) \cdot \bm r_{i j}} t_{c, i j}-\mu \delta_{ij} \}  c^\dagger_{i \sigma} c_{j \sigma}\nonumber  \\
&+ \sum_{i j \sigma} \{e^{i \bm A (t) \cdot \bm r_{i j}} |b|^2  t_{f, i j} + (\epsilon_f + \lambda - \mu )\delta_{ij}   \} f^\dagger_{i \sigma} f_{j \sigma} \nonumber \\
&+\sum_{i j \sigma \sigma^\prime}\{ e^{i \bm A (t) \cdot \bm r_{i j}} b^*V_{i j \sigma \sigma^\prime} c^\dagger_{i \sigma} f_{j \sigma^\prime} + \mathrm{h.c.} \} ,
\label{Model_tdep}
\end{align}
where $\bm A (t) = A\cos \theta_A \cos \omega t \bm e_x + A \sin \theta_A \cos (\omega t - \varphi ) \bm e_y$, $\bm e_x =(1,0,0)$, $\bm e_y =(0,1,0)$, and $\varphi$ is the polarization angle of the laser light. In general, it is difficult to solve a time-dependent quantum many-body problem. However, in the case of time-periodic problem, we can use a useful description with Floquet theory which is known as a theoretical tool for time-periodic systems\cite{Bukov2015}. The model Hamiltonian (\ref{Model_tdep}) is time-periodic and thus we apply Floquet theory \cite{Bukov2015}. 

Floquet theory is based on Floquet's theorem, which is, so to speak, Bloch's theorem for time direction : 
If the Hamiltonian is time-periodic, $\calH(t)=\calH(t+\mathcal{T})$, 
the eigenfunction can be written by a product of an exponential function $e^{-i\epsilon t}$ and a time periodic function $u(t)$. $\epsilon$ is called \textit{quasi-energy} that is defined in the range of $-\pi/\mathcal{T} \leq \epsilon \leq \pi/\mathcal{T}=\omega/2$. 
Then, we can define the effective Hamiltonian, of which eigenvalues are quasi-energies, as
\begin{align}
\calH_{\eff}=\frac{i}{\mathcal{T}}\log U(\mathcal{T}),\label{Heff_def}
\end{align} 
where the time-evolution operator $U(t)=\mathbf{T}\exp \left(-i \int_0^{t} \calH(s)ds \right) $. $\mathbf{T}$ represents the time-ordering operator.
By definition, the effective Hamiltonian has only the information of $t=0,\mathcal{T},2\mathcal{T},\cdots, n\mathcal{T},\cdots$ and gives ``stroboscopic description" of this system. 
In other words, the information of the time range of $n\mathcal{T}<t<(n+1)\mathcal{T}$ is neglected. 
Intuitively, if the time period is short enough, this Hamiltonian tells us the asymptotic behavior of periodically-driven systems.
In fact, the Gibbs ensemble of the effective Hamiltonian describes the nonequilibrium steady states, which are shown to appear with finite lifetime when the frequency is sufficiently high\cite{Abanin2015, Kuwahara2016}. 
Thus we can use the effective Hamiltonian in Floquet theory and also the established techniques in equilibrium physics for understanding the nature of the nonequilibrium steady states of laser-irradiated materials including strongly correlated electron systems.

Though it is difficult to directly calculate the effective Hamiltonian from the definition (\ref{Heff_def}), there are some useful methods to derive the effective Hamiltonian. We adopt a perturbative expansion \cite{Kitagawa2011, Mikami2016} in $1 / \omega$. Using this approach, we can write down the effective Hamiltonian as
\begin{align}
\calH_{\mathrm{eff}} &= \calH_0 + \calH_\mathrm{com} +\calO\left(\omega^{-2}\right), \label{Heff_expansion} \\
\calH_0 &= \calH^{(0)}, \\
\calH_\mathrm{com} &= \sum_{n > 0} \frac{[\calH^{(+n)},\calH^{(-n)}]}{n \omega},
\end{align}
where $\calH^{(n)} =\frac{1}{\mathcal{T}} \int^{\mathcal{T}/2}_{-\mathcal{T}/2}dt \calH(t) e^{-i n \omega t}$. The first term $\calH_0$ is the time-average of the original time-dependent Hamiltonian (\ref{Model_tdep}). The second term $\calH_\mathrm{com}$ represents the second order perturbation process of the $n$-photon absorption $\calH^{(+n)}$ and $n$-photon emission $\calH^{(-n)}$ in off-resonant light.

\section{Derivation of the effective model and the laser-induced effects}

In this section, we derive an effective model of Kondo insulators irradiated by the laser light 
$\bm A (t)$
and show the laser-induced effects which are found in our effective model.
First we calculate the $n$-th Fourier component of the time-dependent Hamiltonian (\ref{Model_tdep}).
For this purpose, we introduce parameters $\rho_{ij} >0 $ and $\theta_A \in [0, 2 \pi )$ as
\begin{align}
\rho_{ij}(\theta_A) e^{i \phi_{i j}(\theta_A)} : = \cos \theta_A \bm e_x \cdot \bm r_{ij}  + \sin \theta_A e^{- i \varphi}\bm e_y  \cdot \bm r_{ij},
\end{align}
and use the Jacobi-Anger expansion
\begin{align}
e^{i z \cos \theta} = \sum^\infty_{n=-\infty} i^n J_n (z) e^{i n \theta},
\end{align}
where $J_n (z)$ represents the $n$-th Bessel function.
Then we obtain the $n$-th Fourier component $\calH^{(n)}$ as
\begin{align}
\calH^{(n)}
&=\sum_{i j \sigma} \{ \calJ^{(n)}_{i j}(A, \theta_A) t_{c, i j}- \mu \delta_{ij} \delta_{n 0} \}  c^\dagger_{i \sigma} c_{j \sigma}\nonumber  \\
&+ \sum_{i j \sigma} \{  \calJ^{(n)}_{i j}(A, \theta_A) |b|^2  t_{f, i j} + (\epsilon_f + \lambda - \mu )\delta_{ij} \delta_{n0}  \} f^\dagger_{i \sigma} f_{j \sigma} \nonumber \\
&+\sum_{i j \sigma \sigma^\prime}\{  \calJ^{(n)}_{i j}(A, \theta_A) b^*V_{i j \sigma \sigma^\prime} c^\dagger_{i \sigma} f_{j \sigma^\prime} \nonumber \\
&\qquad \qquad +  \calJ^{(n)}_{i j}(A, \theta_A) b V_{ j i \sigma^\prime \sigma }^* f^\dagger_{i \sigma} c_{j \sigma^\prime}\},
\end{align}
with $\calJ^{(n)}_{i j}(A, \theta_A)= i^n J_n (A \rho_{ij}(\theta_A)) e^{i n  \phi_{i j}(\theta_A)}$.
We apply this result to the formula (\ref{Heff_expansion}) and then obtain the explicit form of the effective Hamiltonian as 
\begin{align}
\calH_{\mathrm{eff}} &= \calH_0 + \calH_\mathrm{com},\\
\calH_0&
=\sum_{i j \sigma} ( \tilde{t}_{c, i j}- \mu \delta_{ij} )  c^\dagger_{i \sigma} c_{j \sigma} \nonumber \\
&+ \sum_{i j \sigma} \{ |b|^2  \tilde{t}_{f, i j} + (\epsilon_f + \lambda - \mu )\delta_{ij}  \} f^\dagger_{i \sigma} f_{j \sigma} \nonumber \\&+\sum_{i j \sigma \sigma^\prime}\{  b^* \tilde{V}_{i j \sigma \sigma^\prime} c^\dagger_{i \sigma} f_{j \sigma^\prime} + \mathrm{h.c.}\} ,
\end{align}
\begin{widetext}
\begin{align}
\calH_\mathrm{com}&
=\sum_{i j \sigma \sigma^\prime}
\tau_{c, i j \sigma \sigma^\prime}
c^\dagger_{i \sigma} c_{j \sigma^\prime} \nonumber+\sum_{i j \sigma \sigma^\prime} 
|b|^2  \tau_{f, i j \sigma \sigma^\prime}f^\dagger_{i \sigma} f_{j \sigma^\prime}  \\
&+\sum_{i j \sigma \sigma^\prime} \left \{ b^*  \Upsilon_{i j \sigma \sigma^\prime} c^\dagger_{i \sigma} f_{j \sigma^\prime} + \mathrm{h.c.} \right \},
\end{align}
where
\begin{align}
\tilde{t}_{c, i j} &= J_0 (A \rho_{ij}(\theta_A)) t_{c, i j}, \label{hop_c_0}\\
\tilde{t}_{f, i j}&= J_0 (A \rho_{ij}(\theta_A)) t_{f, i j}, \label{hop_f_0} \\
\tilde{V}_{i j \sigma \sigma^\prime}&= J_0 (A \rho_{ij}(\theta_A)) V_{i j \sigma \sigma^\prime} \label{hyb_0},
\end{align}
\begin{align}
\tau_{c, i j \sigma \sigma^\prime} &= 2 i  \sum_k \calJ_{ikj}(A, \theta_A) \frac{t_{c, i k} t_{c, k j} + |b|^2  V_{i k \sigma s} V_{ j k  \sigma^\prime s}^*}{\omega},  \label{hop_c_n}\\
\tau_{f, i j \sigma \sigma^\prime}&= 2 i \sum_k\calJ_{ikj}(A, \theta_A)\frac{|b|^2  t_{f, i k} t_{f, k j} + V^*_{ k i  s\sigma} V_{k j s \sigma^\prime}}{\omega},  \label{hop_f_n}\\
\Upsilon_{i j \sigma \sigma^\prime}&=2 i \sum_k  \calJ_{ikj}(A, \theta_A)\frac{t_{c, i k} V_{k j \sigma \sigma^\prime} + |b|^2  V_{i k \sigma \sigma^\prime} t_{f, k j}}{\omega},  \label{hyb_n}
\end{align}
\begin{align}
\calJ_{ikj}(A, \theta_A) &= \sum_{n >0}\frac{ (-1)^n J_n (A \rho_{ik}(\theta_A)) J_n (A \rho_{kj}(\theta_A))}{n} \sin \left[ n (\phi_{i k}(\theta_A)- \phi_{k j}(\theta_A)  )\right].
\end{align}
\end{widetext}

In this effective model, we find two types of effect induced by the laser light. 

i) \textit{Dynamical localization}: 
Looking at the zero-th order term $\calH_0$, we find that 
the amplitudes of hopping in (\ref{hop_c_0}), (\ref{hop_f_0}) and hybridization  in (\ref{hyb_0}) are renormalized by zero-th Bessel function.
Thus the amplitudes decrease with increasing the laser intensity. 
This effect corresponds to the phenomenon called dynamical localization \cite{Dunlap1986}, 
which causes freezing of the motion of electrons when the system is irradiated by a strong electric field \cite{Ishikawa2014}.
The important point in our study is that this affects only off-site terms ($i \neq j$) since the Peierls phase is equal to zero in on-site terms ($i = j$).

ii) \textit{Laser-induced hopping and hybridization}: 
In the commutator term $\calH_\mathrm{com}$, we find the new hopping terms (\ref{hop_c_n}), (\ref{hop_f_n}) and hybridization term (\ref{hyb_n}) which do not exist in the original Hamiltonian (\ref{Model_PAM}). 
They depend not only on the position but also on the spins of electrons.
Thus these terms can be regarded as an effective spin-orbit coupling.
They give a drastic effect on the renormalized band structure, such as a level splitting.
We note that these terms can break the time-reversal symmetry depending on the polarization of the laser field. In such a case they can be interpreted as an effective magnetic field. 

The important point is that these two effects are derived from the quite general model which describes heavy fermion systems.
Thus these effects are expected to appear generically in heavy fermion systems. 

In the following sections, we demonstrate the physical consequences of these effects with calculations of concrete models.
In Sec. IV, we show that the dynamical localization affects the Kondo effect. We find that the Kondo effects are enhanced or suppressed by laser light depending on whether the hybridization is on-site or not. 
In Sec. V, we demonstrate that various laser-induced topological phase transitions are realized in a certain type of Kondo insulators. 

\section{Controlling Kondo effect}

\subsection{On-site model and off-site model}

\begin{figure}
\includegraphics[width=7cm]{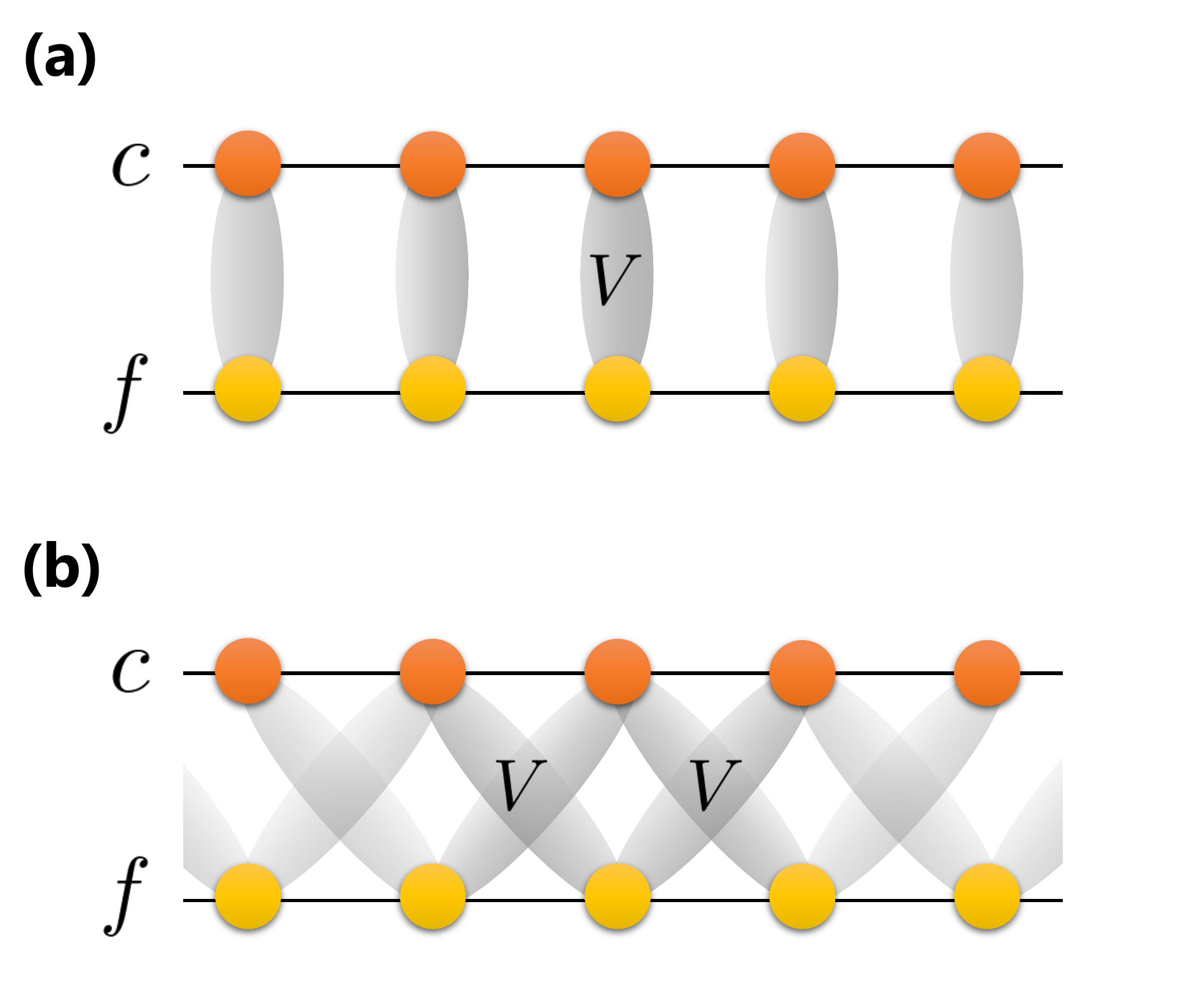}
\caption{Schematic picture of $c$-$f$ hybridization. 
(a) on-site model: The $c$- and $f$- electrons hybridize only at the same site
(b) off-site model: The $c$- and $f$- electrons hybridize only at the nearest-neighbor sites. There is no on-site hybridization.
Note that, although this figure treats one-dimensional systems for simplicity, we consider three-dimensional systems in our study,}
\label{fig_hyb}
\end{figure}

In order to demonstrate the effect of laser light on the Kondo effect, which has been proposed in Sec. III, 
we introduce two simple models, which show the qualitatively opposite behavior under the laser light as we see below. One is an on-site model $\calH_\mathrm{on}$, which has only an on-site hybridization term. 
The other is an off-site model $\calH_\mathrm{off}$, in which the localized electrons can hybridize to the conducting orbit only at the nearest neighbor sites. The structures of hybridization are schematically shown in Fig. \ref{fig_hyb}. For simplicity, we assume cubic symmetry and then the Hamiltonian reads
\begin{align}
\calH_\mathrm{on/off}&= \calH^{(c)} + \calH^{(f)} + \calH_\mathrm{on/off}^\mathrm{(hyb)} +\calH^\mathrm{(int)},
\end{align}
\begin{align}
\calH^{(c)} &=\sum_{\bm k \sigma} (\epsilon_c (\bm k)-\mu) c^\dagger_{\bm k \sigma} c_{\bm k \sigma},  \\
\calH^{(f)}  &=\sum_{\bm k \sigma} (\epsilon_f (\bm k)-\mu) f^\dagger_{\bm k \sigma} f_{\bm k \sigma}, \\
\calH^{(\mathrm{int})} &=U \sum_i n_{i \uparrow}^{(f)} n_{i \downarrow}^{(f)}, \\
\calH_\mathrm{on}^\mathrm{(hyb)}&=\sum_{\bm k \sigma} \{ V  c^\dagger_{\bm k \sigma} f_{\bm k \sigma} + \mathrm{h.c.} \},\label{on_lin_hyb_term} \\
\calH_\mathrm{off}^\mathrm{(hyb)}&=\sum_{\bm k \sigma} \{ V (\bm k) c^\dagger_{\bm k \sigma} f_{\bm k \sigma} + \mathrm{h.c.} \},\label{off_lin_hyb_term}
\end{align}
with
\begin{align}
\epsilon_c ({\bm k})&=-2 t_c(\cos k_x+\cos k_y +\cos k_z), \label{ec} \\ 
\epsilon_f ({\bm k})&=\epsilon_f-2 t_f(\cos k_x+\cos k_y +\cos k_z), \label{ef} \\
V (\bm k) &= i V (\sin k_x+\sin k_y +\sin k_z).
\end{align}
Here $c_{\bm k\sigma}$ and $f_{\bm k\sigma}$ are the momentum representations of the annihilation operators.
Note that the parity of the conduction electron is fixed in the models, i.e. the on-site (off-site) model corresponds to odd (even) parity,
since the parity of the localized electrons ($f$-orbit) is odd. 
Therefore most of heavy fermion materials ($s$-$f$ or $d$-$f$ hybridization) are represented by $\calH_\mathrm{off}$ and the heavy fermion systems with $p$-$f$ hybridization are represented by $\calH_\mathrm{on}$.

\floatsetup[figure]{style=plain,subcapbesideposition=top}
\begin{figure*}
\sidesubfloat[]{\includegraphics[width=8cm]{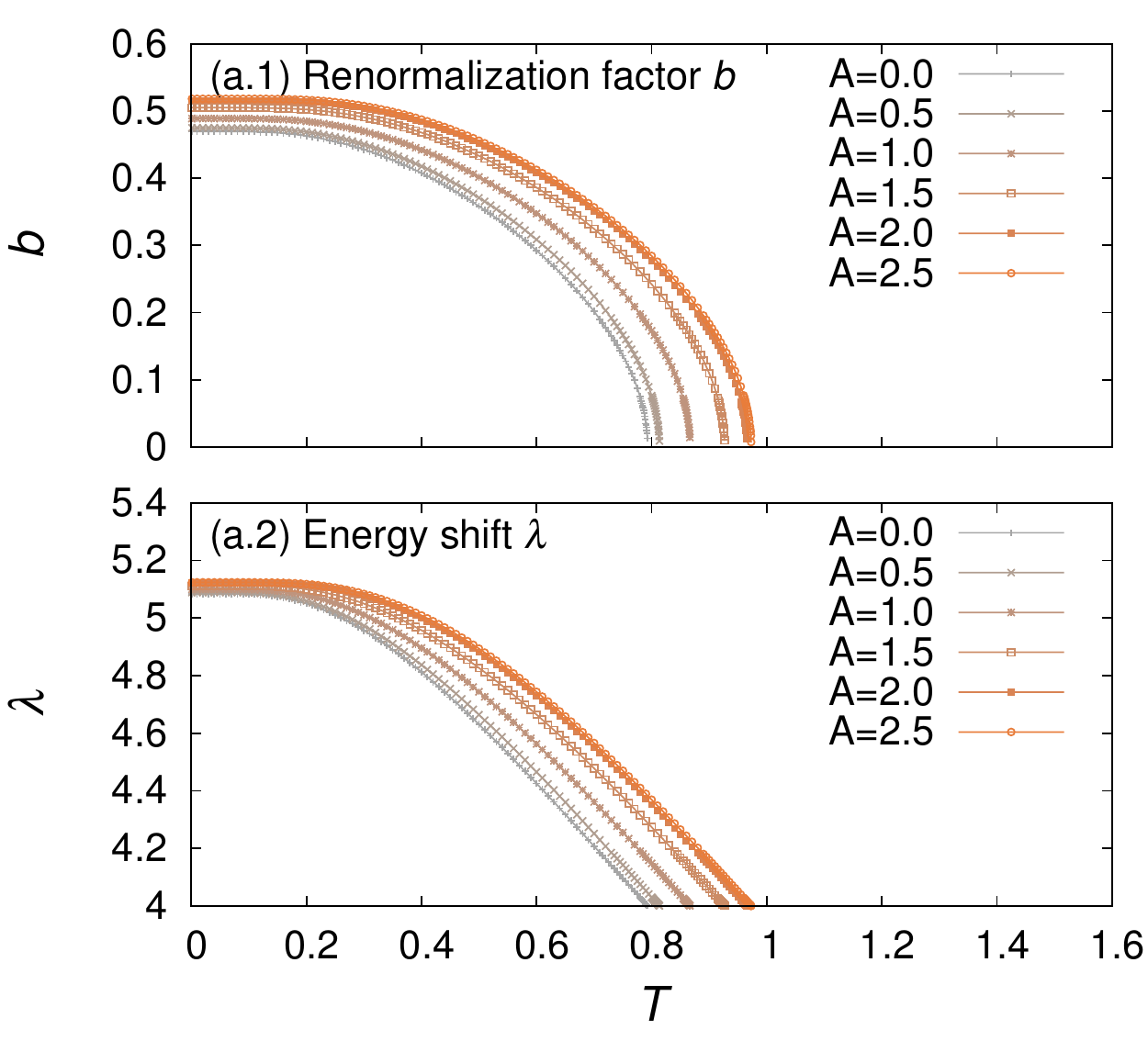} } \ \ \ \
\sidesubfloat[]{\includegraphics[width=8cm]{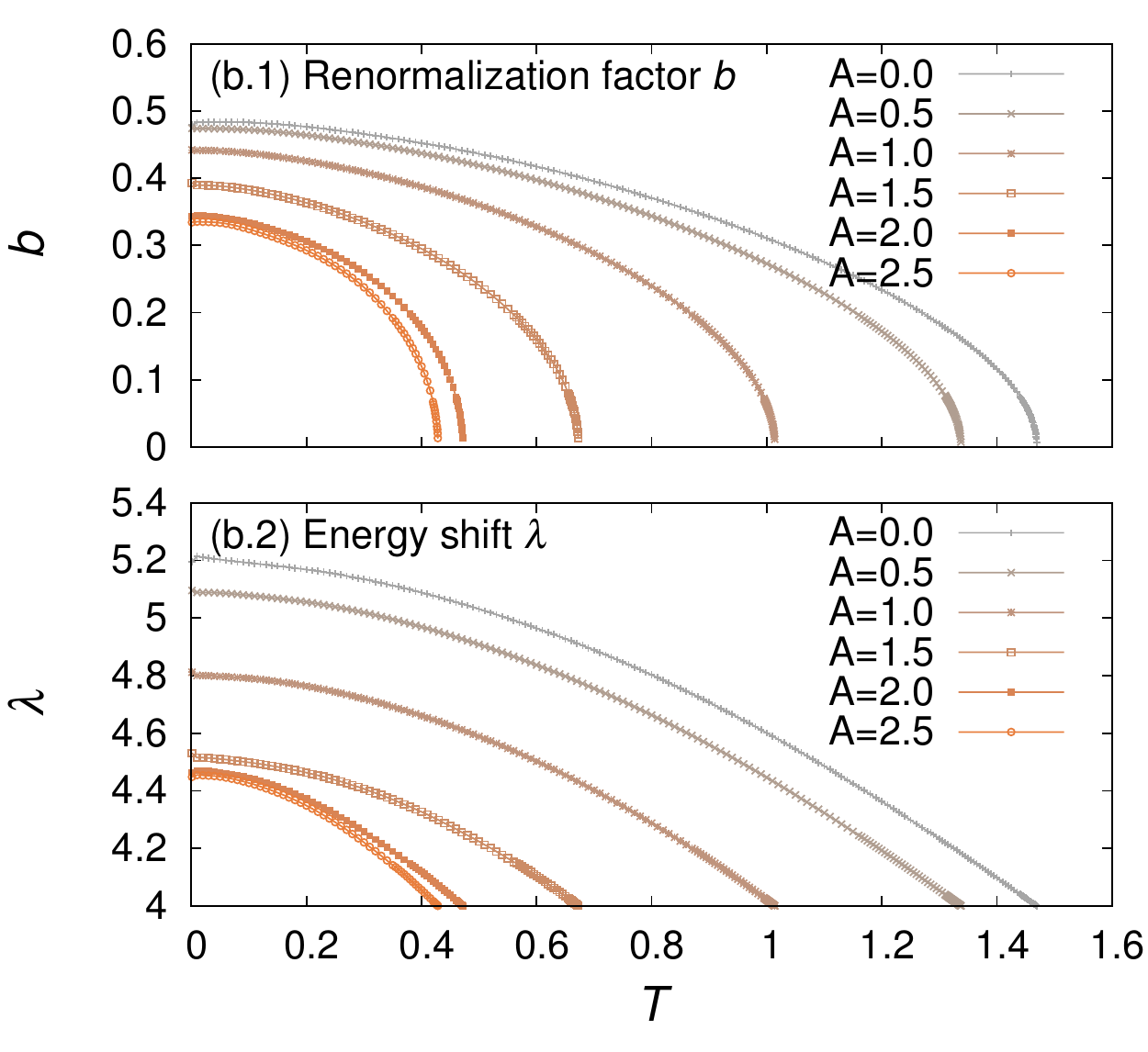} }\\%
\caption{Numerical solutions of the self-consistent equations for (a) the on-site model and (b) the off-site model with the linearly polarized laser light. We show the temperature dependence of (a.1) the renormalization factor $b$ and (a.2) the energy shift $\lambda$ for the on-site model. The same things for the off-site model are shown in Fig. (b.1) and (b.2) respectively. 
Here $A_x=A_y=A$ and we use the parameters as $t_c=1$, $t_f=-0.2$, $V=3$, $\epsilon_f=-4$ and $\omega=12.5$.}\label{lin}
\end{figure*}

These models correspond to the specific cases of the model discussed in Sec.III. 
Thus we can apply the results in Sec.III to these models.
With linearly polarized laser light ($\varphi = 0$), we obtain the effective models as
\begin{align}
\calH^{(\mathrm{lin})}_\mathrm{eff, on/off}&= \tilde{\calH}^{(c)} + \tilde{\calH}^{(f)} + \tilde{\calH}_\mathrm{on/off}^\mathrm{(hyb, lin)} \label{Heff_onoff_lin},
\end{align}
\begin{align}
\tilde{\calH}^{(c)} &=\sum_{\bm k \sigma} (\tilde{\epsilon_c} (\bm k)-\mu) c^\dagger_{\bm k \sigma} c_{\bm k \sigma},  \\
\tilde{\calH}^{(f)}  &=\sum_{\bm k \sigma} (\tilde{\epsilon_f} (\bm k, |b|)+\lambda-\mu) f^\dagger_{\bm k \sigma} f_{\bm k \sigma}, \\
\tilde{\calH}_\mathrm{on}^\mathrm{(hyb, lin)}&=\sum_{\bm k \sigma} \{ V b^* c^\dagger_{\bm k \sigma} f_{\bm k \sigma} + \mathrm{h.c.} \}, \\
\tilde{\calH}_\mathrm{off}^\mathrm{(hyb, lin)}&=\sum_{\bm k \sigma} \{ \tilde{V} (\bm k)  b^*  c^\dagger_{\bm k \sigma} f_{\bm k \sigma} + \mathrm{h.c.} \},
\end{align}
with
\begin{align}
\tilde{\epsilon}_c ({\bm k})&=-2 t_c(J_0(A_x) \cos k_x+ J_0(A_y)\cos k_y +\cos k_z), \label{tilde_ec}
\end{align}
\begin{align}
\tilde{\epsilon}_f ({\bm k})&=\epsilon_f - 2 t_f |b|^2 (J_0(A_x) \cos k_x+ J_0(A_y)\cos k_y +\cos k_z), \label{tilde_ef}
\end{align}
\begin{align}
\tilde{V} (\bm k) &= iV (J_0(A_x) \sin k_x+J_0(A_y) \sin k_y +\sin k_z). \label{tilde_v}
\end{align}
Here we define $A_x$ and $A_y$ as $(A_x, A_y) = A ( \cos \theta_A,  \sin \theta_A)$. 
The important difference between the on-site and off-site models is the existence/absence of the dynamical localization effect in the hybridization.
Note that this effective Hamiltonian includes only the zero-th order term $\calH_0$. Namely, the commutator contribution, $\calH_\mathrm{com}$, vanishes. Thus there are only the effects of the dynamical localization which can be seen in eqs. (\ref{tilde_ec}) - (\ref{tilde_v}).
On the other hand, in the case of the circularly polarized laser light  ($\varphi = - \pi/2$), the effective Hamiltonian includes the commutator contribution, $\calH_\mathrm{com}$, as 
\begin{align}
\calH^{(\mathrm{cir})}_\mathrm{eff, on/off}&= \tilde{\calH}^{(c)} + \tilde{\calH}^{(f)} + \tilde{\calH}_\mathrm{on/off}^\mathrm{(hyb, cir)},\label{Heff_onoff_cir}
\end{align}
with
\begin{align}
\tilde{\calH}_\mathrm{on}^\mathrm{(hyb, cir)}&=\tilde{\calH}_\mathrm{on}^\mathrm{(hyb, lin)}=\sum_{\bm k \sigma} \{ V b^* c^\dagger_{\bm k \sigma} f_{\bm k \sigma} + \mathrm{h.c.} \}, \label{hyb_on_cir} \\
\tilde{\calH}_\mathrm{off}^\mathrm{(hyb, cir)}&=\sum_{\bm k \sigma} \{ (\tilde{V} (\bm k)+ \Upsilon (\bm k))  b^*  c^\dagger_{\bm k \sigma} f_{\bm k \sigma} + \mathrm{h.c.} \}.
\end{align}
Here we have defined
\begin{align}
\Upsilon (\bm k)&= -\frac{4 (t_c- |b|^2 t_f) V}{\omega} \calJ(A_x, A_y) \sin (k_x - k_y ),\\
\calJ(A_x, A_y) &=\sum_{m=0} \frac{(-1)^m J_{2m+1} (A_x) J_{2m+1} (A_y)}{2m+1} \label{calJ}.
\end{align}
$\Upsilon (\bm k)$ in $\tilde{\calH}_\mathrm{on/off}^\mathrm{(hyb,cir)}$ comes from $\calH_\mathrm{com}$. 
This term corresponds to the laser-induced hybridization which does not exist in the original model and connects a site with its next-nearest-neighboring sites. 
Moreover, this term gives rise to the breaking of time-reversal symmetry, which is inherent in the circularly polarized laser light. As we see below, this effect only gives a small contribution to the Kondo effect. However, this kind of laser-induced hopping and hybridization plays a crucial role for the topological properties.
This point is discussed in Sec. V.  

\subsection{Numerical results}
\subsubsection{Linearly polarized laser light}

Following the methods shown in Sec. II A, we numerically calculate the renormalization factor $b$ and the energy shift of the $f$-electron level $\lambda$ self-consistently for the effective models (\ref{Heff_onoff_lin}) and (\ref{Heff_onoff_cir}). Then, based on the results, we discuss the effect of laser light on the Kondo effect. We note here that in the slave boson approach, the fomation of Kondo singlets is well described in terms of the renormalization factor $b$, although the crossover behavior at finite temperatures appears as a phase transition with the order parameter $b$.  Although this transition is artificial, it is known that the transition temperature corresponds to the Kondo temperature \cite{Coleman1983, Coleman1987, Newns1987}, which is a characteristic temperature where the Kondo effect takes place. Below the Kondo temperature, we can correctly describe the formation of the Kondo singlets and the resulting insulating behavior characteristic of the Kondo insulator.

First we consider the case of the linearly polarized laser light. The results for the on-site model are shown in Fig. \ref{lin}(a).
From Fig. \ref{lin}(a), we find that the Kondo temperature is enhanced with increasing the intensity of laser light. Thus the Kondo effect is enhanced in the on-site model. On the other hand, we find that the Kondo temperature decreases as we increase the intensity of laser light in Fig. \ref{lin}(b) for the off-site model.
This means that the Kondo effect is suppressed in the off-site model.
These results suggest that whether the Kondo effect is enhanced or suppressed depends on the spatial structure of the hybridization. Only in the off-site model, the hybridization terms couple to the electromagnetic fields and then the dynamical localization takes place in eq.(\ref{off_lin_hyb_term}). As seen in eq. (\ref{on_lin_hyb_term}), the on-site model does not show the dynamical localization in the hybridization. This difference plays a crucial role in generating the opposite behavior of the Kondo effect in those models. The origin of this behavior is discussed in Sec. IV C.

\begin{figure}[t]
\includegraphics[width=8cm]{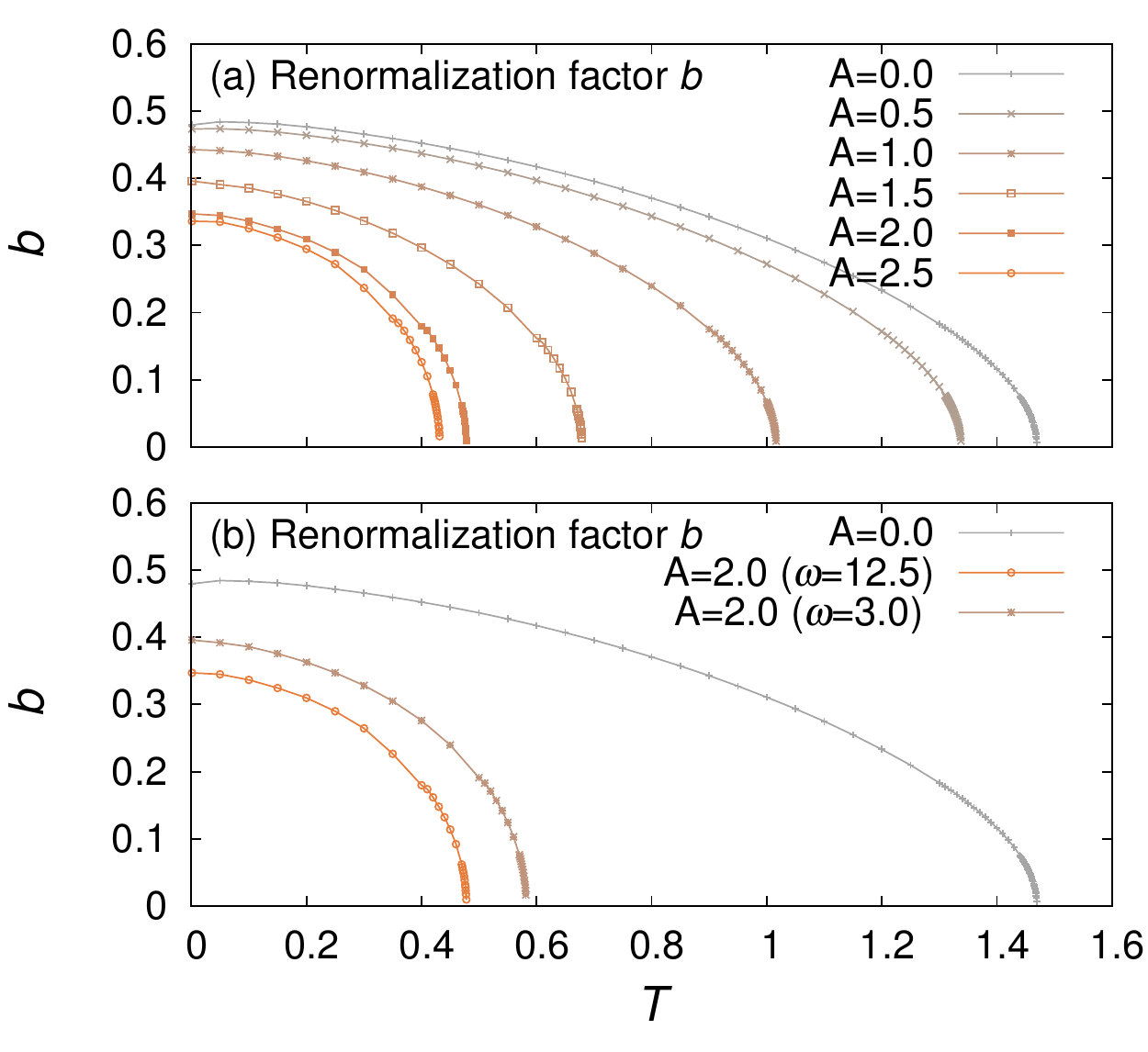}
\caption{Numerical solutions of the self-consistent equations for the off-site model with the circularly polarized laser light. In Fig. (a), we show the temperature dependence of the renormalization factor $b$. Here $A_x=A_y=A$ and we use the parameters as $t_c=1$, $t_f=-0.2$, $V=3$, $\epsilon_f=-4$ and $\omega=12.5$. In Fig.(b), to clarify the effect of laser-induced hybridization, we compare the results of $\omega =3.0$ and $\omega = 12.5$ at the same intensity ($A=2.0$). All the parameters except for the frequency are set same as Fig. (a). }
\label{off_cir}
\end{figure}

\subsubsection{Circularly polarized laser light}
Next we consider the circularly polarized laser light.
For the on-site model, the effective model with circularly polarized laser light is the same as the one with linearly polarized laser light, shown in eq. (\ref{hyb_on_cir}). Thus we study only the off-site model in this subsection.

The results for the circularly polarized laser are shown in Fig. \ref{off_cir}.
From Fig \ref{off_cir} (a), we find that the Kondo temperature is suppressed by increasing the laser intensity in the same manner as in the case of the linearly polarized laser. The behavior is quite similar to that in Fig \ref{lin}(b.1) and thus the effect of the laser-induced hybridization, which appears only when the laser is circularly polarized, is difficult to find out. To clarify the reason, we examine the low-frequency regime ($\omega = 3.0$), in which our high frequency expansion could still predict the qualitative tendency. The results are shown in Fig. \ref{off_cir}(b). 
We find that the Kondo temperature in the low frequency regime ($\omega = 3.0$) is higher than the one in the high frequency regime ($\omega = 12.5$) and
thus confirm that the laser-induced hybridization indeed enhances the Kondo effect. Therefore, the reason why this effect is difficult to find out in the high frequency regime ($\omega=12.5$) is that the extra hybridization $\Upsilon (\bm k)$ is much smaller than the zero-th order term $\calH_0$ in the high frequency expansion and then does not give an important effect on the Kondo effect.

\subsection{Enhancement/Suppression of Kondo effect}

\begin{figure}[t]
\sidesubfloat[]{\includegraphics[width=8cm]{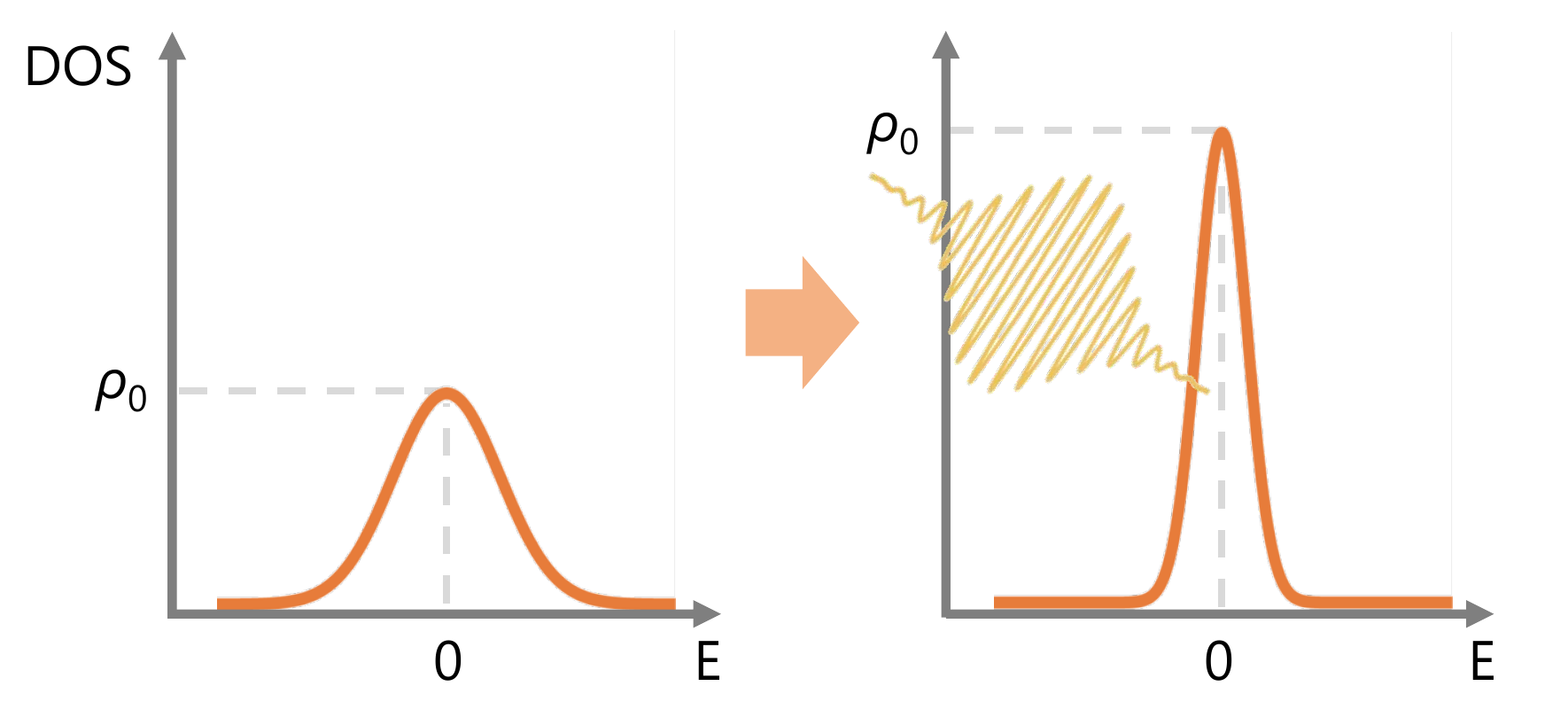} }\\
\sidesubfloat[]{\includegraphics[width=6.5cm]{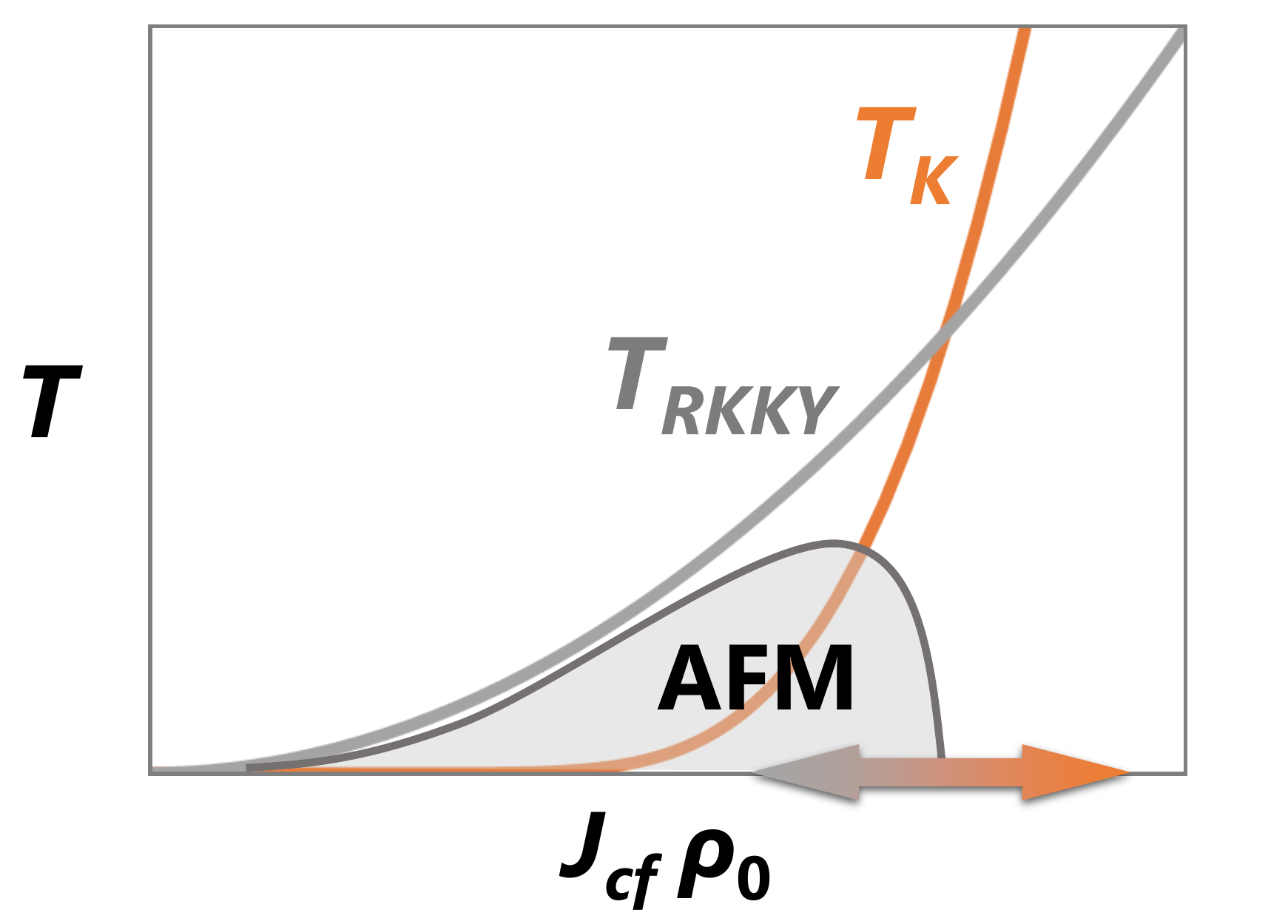} }
\caption{(a) The schematic picture of the enhancement of the density of states near the Fermi energy $\rho_0$. 
It is the origin of the enhancement of the Kondo effect in the on-site model.
(b) Schematic Doniach phase diagram. 
Changing the Kondo coupling $J_{cf}$ or the density of states $\rho_0$ by some perturbations, e.g. pressure, we can induces the quantum phase transition. AFM represents an antiferromagnetic phase.}
\label{DOS_Doniach}
\end{figure}

In order to understand the origin of the qualitatively opposite behavior of the Kondo effect in the on-site and off-site model, we give a rough estimation of the Kondo temperature in this subsection.
In the case of infinite $U$, it is known that the Kondo temperature is approximately given as \cite{Coleman1983}
\begin{align}
T_K \simeq D \exp \left( \frac{\epsilon_f - \mu}{2 \rho_0 V^2} \right). \label{TK}
\end{align}
Here $D$ is the band width of the conduction electrons and $\rho_0$ is the density of states at the Fermi energy.
We incorporate the effect of the laser light in this formula.
The effective form of the conduction band (\ref{ec}) shows that the hopping amplitude is renormalized by the zero-th Bessel function, and then
we treat this effect by replacing $t_c$ with $\tilde{t_c} = J_0 (A) t_c$.
Here $A$ is the amplitude of the laser light.
Considering $D \sim t$ and $\rho_0 \sim D^{-1}$, we obtain the Kondo temperature of the laser-irradiated model as 
\begin{align}
\tilde{T}^\mathrm{(on)}_K (A) \simeq J_0 (A) D \exp \left( J_0(A) \frac{\epsilon_f - \mu}{2 \rho_0 V^2} \right). \label{TK_on}
\end{align}
Since $\epsilon_f - \mu$ is negative and thus this formula implies that the Kondo temperature increases in the low intensity region (i.e. small $A$). This estimation shows a qualitative agreement with the numerical results in Sec. IV B. 
This agreement suggests that the dynamical localization plays the key role in this enhancement of Kondo effect. The dynamical localization makes the band width narrower, thereby enhancing the density of states near the Fermi surface (see Fig. \ref{DOS_Doniach}(a)). This enhancement of the density of states helps the formation of Kondo singlet.

On the other hand, in the off-site model, we have to additionally consider the dynamical localization of hybridization and thus we treat this effect by replacing $V$ with $\tilde{V} = J_0 (A) V$. Then we obtain
\begin{align}
\tilde{T}^\mathrm{(off)}_K (A) \simeq J_0 (A) D \exp \left( (J_0(A))^{-1} \frac{\epsilon_f - \mu}{2 \rho_0 V^2} \right). \label{TK_off}
\end{align}
This formula shows that the Kondo temperature of the laser-irradiated off-site model decreases in the low intensity region. This is a qualitatively different behavior from the on-site model. The origin of th suppression is the dynamical localization effect in the $c$-$f$ hybridization. 

Before closing this section, we remark on a possibility to realize laser-induced quantum phase transitions.
In heavy fermion systems, the interplay of the Kondo effect and the Ruderman-Kittel-Kasuya-Yosida (RKKY) interaction plays a crucial role.
Their competition gives various quantum phases, including the magnetically ordered phases, and the resulting phases are well described by the Doniach phase diagram \cite{Coleman_book} shown in Fig.\ref{DOS_Doniach}(b). Moreover, some perturbations to the system, e.g. pressure, possibly drive the quantum phase transitions, which are indeed observed in various materials in heavy fermion systems. 
We have shown that the Kondo effect can be enhanced or suppressed by laser light. Therefore, we can make either the Kondo effect or the RKKY interaction dominant. In the case that the Kondo effect is suppressed, the RKKY interaction becomes dominant and the system can realize a magnetically ordered phase.
Applying the laser light effectively changes the Kondo coupling $J_{cf}$ and the density of states $\rho_0$. Then the system goes beyond the quantum critical point and shows a transition to a magnetic phase (typically, an antiferromagnetic phase) in Doniach phase diagram in Fig.\ref{DOS_Doniach}(b). This laser-induced quantum phase transition opens a new way to dynamically control the magnetic phases by laser light.

\section{Controlling topological phases}

\subsection{Model of topological Kondo insulators}

Some kinds of Kondo insulators show topological phases thanks to their peculiar hybridization structures.
They are called topological Kondo insulators \cite{Dzero2010, Dzero2015, Hagiwara2016}, which are gathering the great attention as the analogue of topological insulators in strongly correlated electron systems. In this section, we demonstrate that the laser light changes the topological properties of the topological Kondo insulators. We use a variant of the periodic Anderson model which has off-site and (pseudo)spin-dependent hybridization terms, which is already introduced as a model of topological Kondo insulators in the previous works \cite{Takimoto2011, Dzero2015}. The model Hamiltonian reads
\begin{align}
\calH&=\sum_{\bm k \sigma} (\epsilon_c (\bm k)-\mu) c^\dagger_{\bm k \sigma} c_{\bm k \sigma}
+\sum_{\bm k \sigma} (\epsilon_f (\bm k)-\mu) f^\dagger_{\bm k \sigma} f_{\bm k \sigma} \nonumber \\
&+\sum_{\bm k \sigma \sigma'} \{ (\bm V(\bm k)  \cdot \bm \sigma)_{\sigma \sigma'} c^\dagger_{\bm k \sigma} f_{\bm k \sigma'} + \mathrm{h.c.} \}+U \sum_i n_{i \uparrow}^{(f)} n_{i \downarrow}^{(f)}, \label{Model_TKI}
\end{align}
with
\begin{align}
\bm V(\bm k)&= V( a_1 \sin k_x, a_2 \sin k_y, a_3 \sin k_z).
\end{align}
Here $\sigma, \sigma^\prime (= \uparrow, \downarrow)$ stands for the (pseudo)spin and $\bm \sigma$ are Pauli matrices. In this study, the parameters are set as $a_1 = 2$, $a_2 = -2$ and $a_3 = 4$. $\epsilon_c ({\bm k})$ and $\epsilon_f ({\bm k})$ have already been defined in eqs. (\ref{ec}) and (\ref{ef}).  Since $\epsilon_c ({\bm k})$ and $\epsilon_f ({\bm k})$ are parity even and $\bm V(\bm k)$ is parity odd, this model has time-reversal symmetry, and thus topological states are characterized by a $Z_2$ topological invariant \cite{Moore2007, FuPRL2007, Dzero2010}. With slave boson mean-fields $b$ and $\lambda$, we can write down the topological invariants $\nu_{\mathrm{STI}}$ and $\nu^\alpha_{\mathrm{WTI}}$ ($\alpha = x, y, z$) for the renormalized band structure in a simplified form \cite{FuPRB2007} as
\begin{align}
(-1)^{\nu_{\mathrm{STI}}}&=\prod_m \delta_m, \label{nu_STI} \\
(-1)^{\nu^\alpha_{\mathrm{WTI}}}&=\prod_m \delta_m \Big|_{(\bm k^*_m)_\alpha =0} \label{nu_WTI},
\end{align}
where $\bm k_m$ represents the time-reversal invariant momenta (TRIM) in three dimensional Brillouin zone, $\delta_m=\mathrm{sgn}(\epsilon_c(\bm k_m)-\bar{\epsilon}_f(\bm k_m, |b|^2)-\lambda)$ is the parity eigenvalue on each TRIM, and $\bar{\epsilon}_f ({\bm k})=\epsilon_f-2 t_f |b|^2 (\cos k_x+\cos k_y +\cos k_z)$.
Note that $\nu_{\mathrm{STI}} = 1$ ($\nu^\alpha_{\mathrm{WTI}}= 1$) means the system is a strong (weak) topological insulator. 

\subsection{Linearly polarized laser light}

\begin{figure}[t]
\includegraphics[width=8cm]{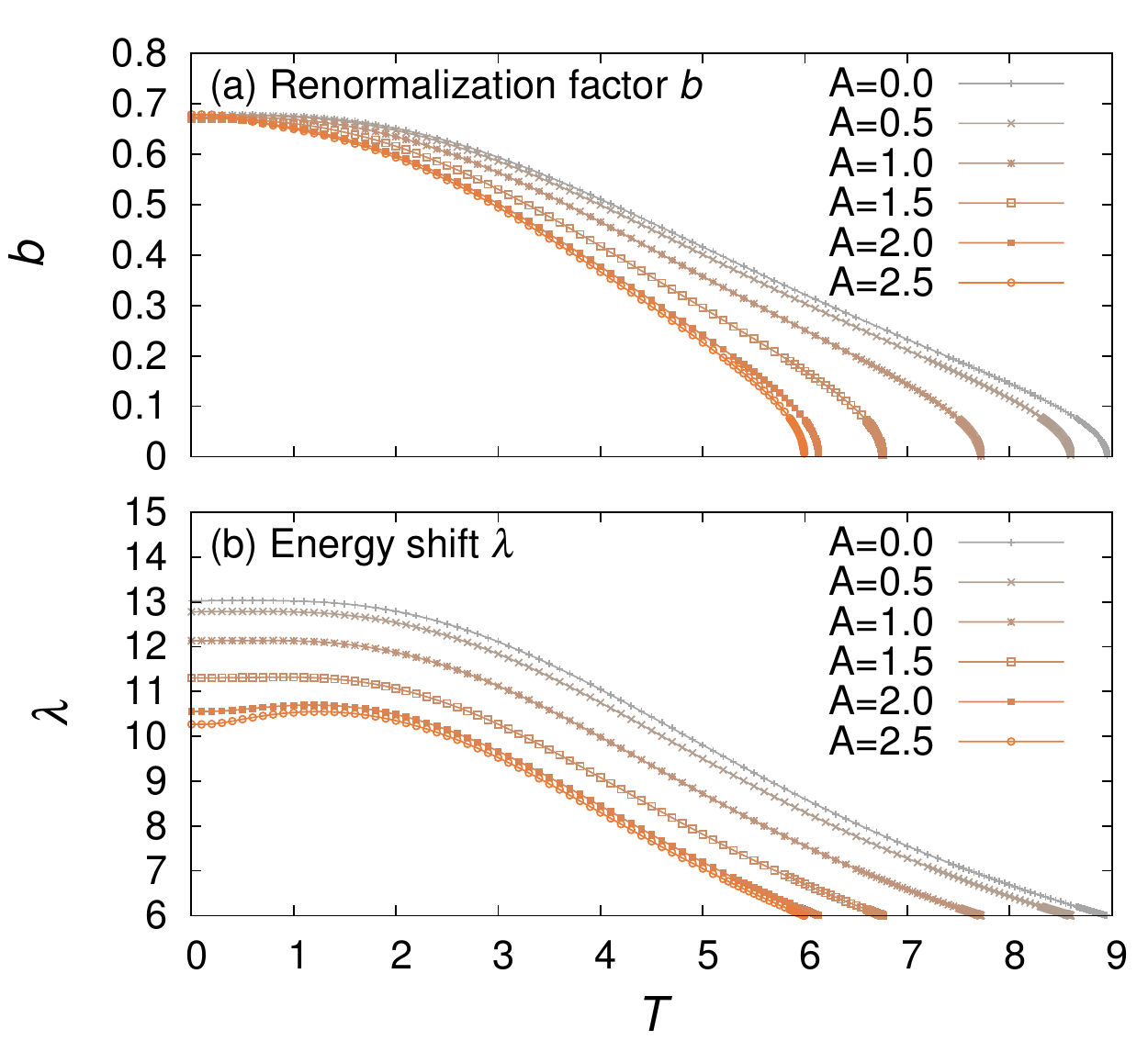} 
\caption{Numerical solutions of the self-consistent equations for the model of  topological Kondo insulator with the linearly polarized laser light. In Fig.  (a) and Fig.  (b), we show the temperature dependence of the renormalization factor $b$ and the energy shift $\lambda$ respectively. 
Here $A_x=A_y=A$ and we use the parameters as $t_c=1$, $t_f=-0.2$, $V=3$, $\epsilon_f=-6$ and $\omega=12.5$.}
\label{TKI_lin}
\end{figure}

First we consider the application of the linearly polarized light. 
The model introduced above corresponds to a specific case of the model which we discussed in Sec. III. 
Thus we can apply the results in Sec. III to this model. 
With linearly polarized laser light  ($\varphi = 0$), 
we obtain the effective model as
\begin{align}
\calH_\eff&=\sum_{\bm k \sigma} \{ \tilde{\epsilon_c} (\bm k)-\mu \} c^\dagger_{\bm k \sigma} c_{\bm k \sigma} \nonumber \\
&+\sum_{\bm k \sigma} \{ \tilde{\epsilon}_f ({\bm k}) + \lambda -\mu\} f^\dagger_{\bm k \sigma} f_{\bm k \sigma} \nonumber \\
&+\sum_{\bm k \sigma \sigma^\prime} \{b^* (\tilde{\bm V} (\bm k) \cdot \bm \sigma)_{\sigma \sigma^\prime} c^\dagger_{\bm k \sigma} f_{\bm k \sigma^\prime} + \mathrm{h.c.} \},\label{Heff_TKI_lin}
\end{align}with
\begin{align}
\widetilde{\bm V}(\bm k)&=V( a_1 J_0(A_x) \sin k_x, a_2 J_0(A_y) \sin k_y, a_3 \sin k_z).
\end{align}
$\tilde{\epsilon_c} (\bm k)$ and $\tilde{\epsilon}_f ({\bm k})$ have already been defined in eqs. (\ref{tilde_ec}) and (\ref{tilde_ef}).
As with the case of on/off-site model in Sec. IV, this effective Hamiltonian includes only the contributions coming from the zero-th order term $\calH_0$. Thus there is only the effect of the dynamical localization of hopping and hybridization. We discuss how these dynamical localization effects change the topological nature of this system.

\begin{figure}[t]
\includegraphics[width=7.6cm]{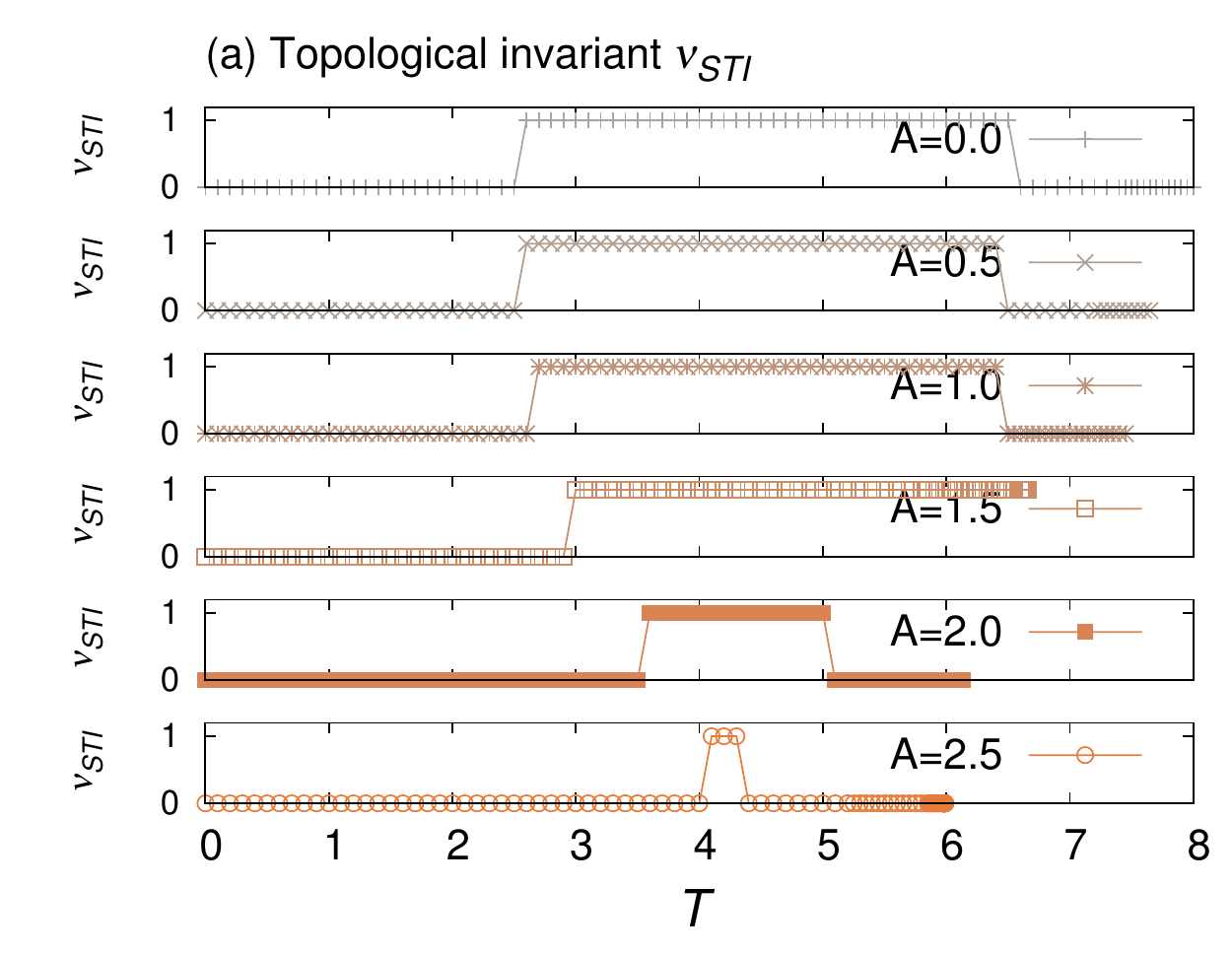} 
\includegraphics[width=7.6cm]{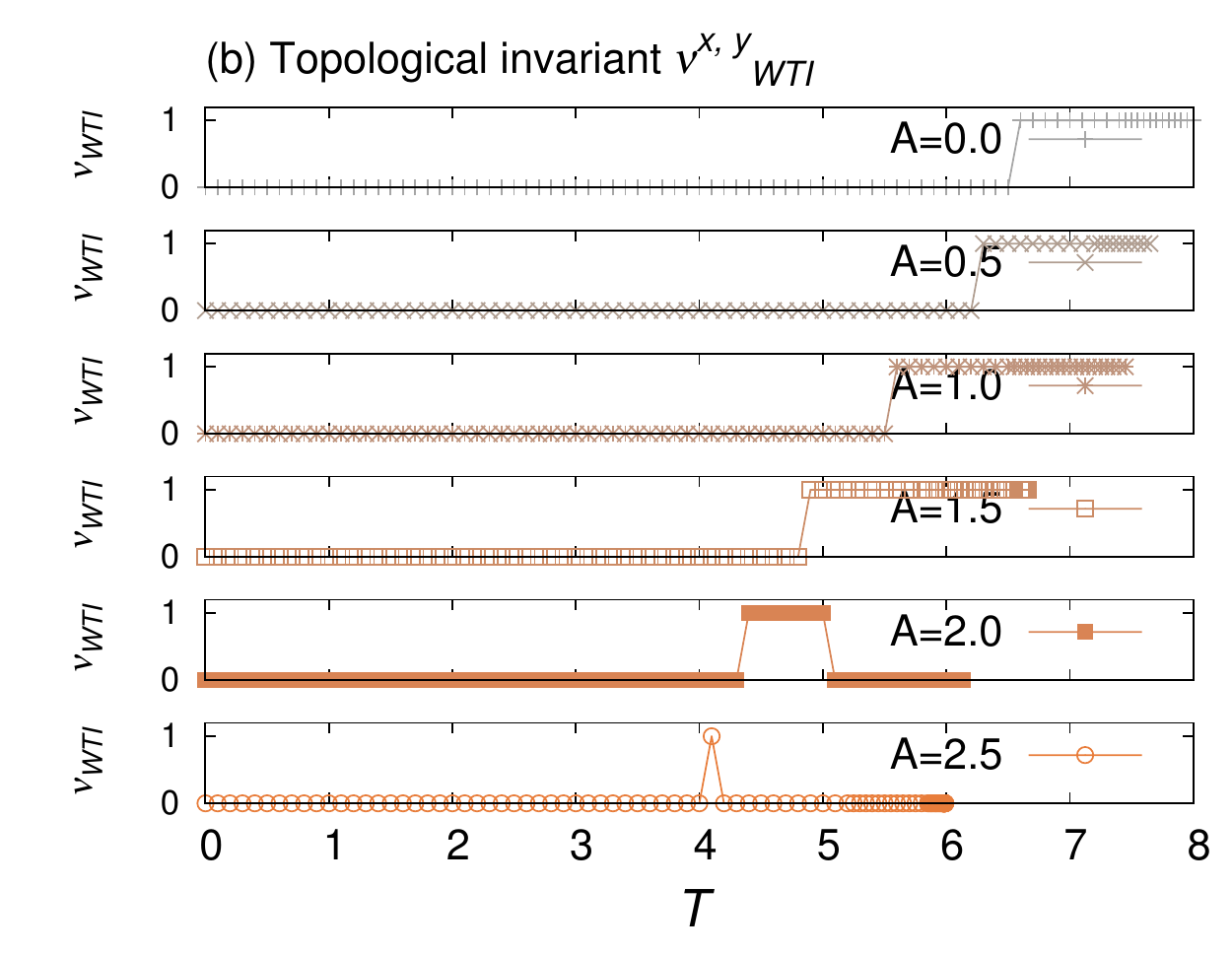}
\includegraphics[width=7.6cm]{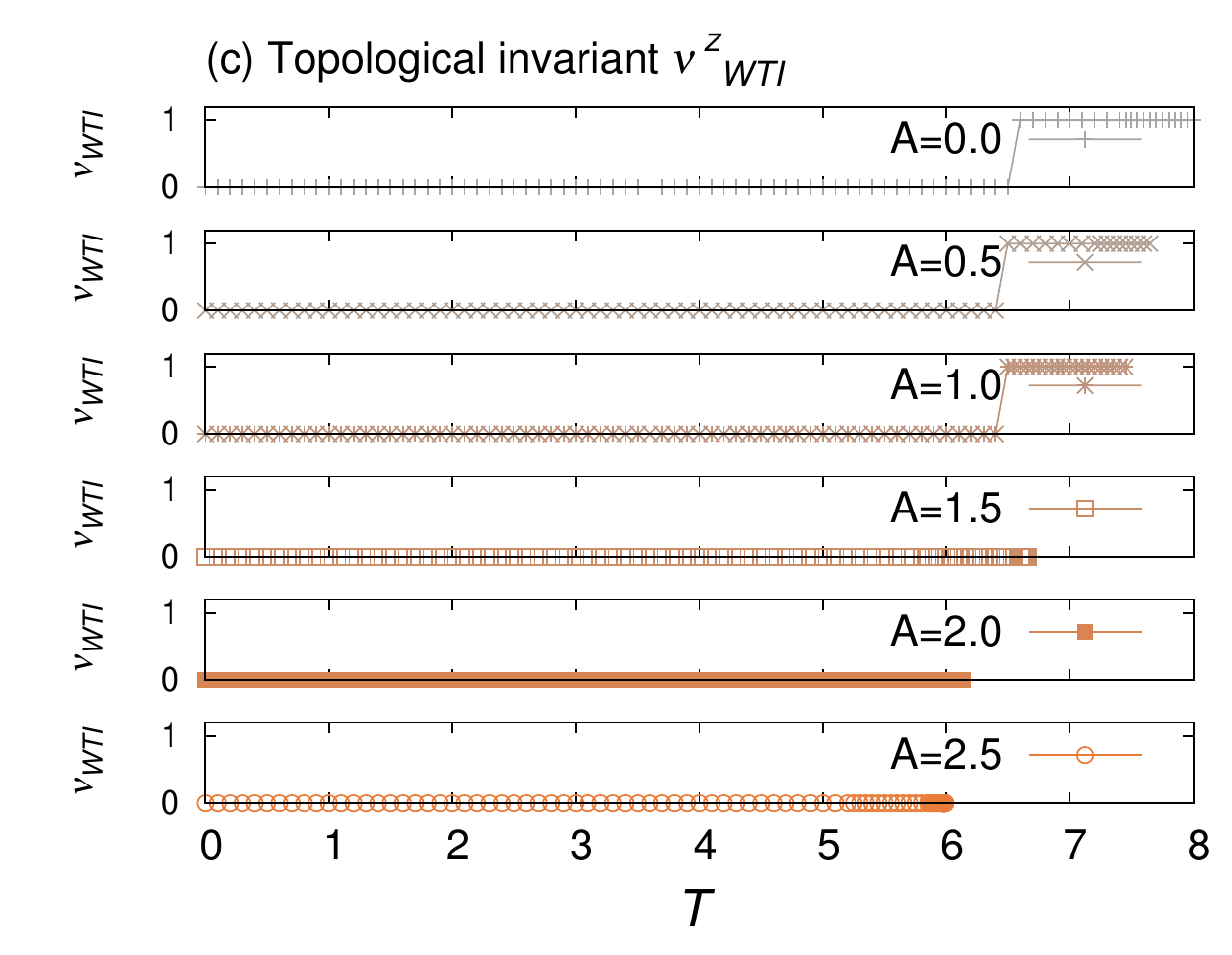}
\caption{Topological invariants (a) $\nu_\mathrm{STI}$, (b) $\nu^{x,y}_\mathrm{WTI}$ and, (c) $\nu^{z}_\mathrm{WTI}$.
Above the Kondo temperature, the topological invariants are not defined since the system is a paramagnetic metal. Here we use the parameters as $t_c=1$, $t_f=-0.2$, $V=3$, $\epsilon_f=-4$ and $\omega=12.5$.}
\label{TKI_lin_top}
\end{figure}

To discuss the topological properties, we use the topological invariants. 
In the effective model (\ref{Heff_TKI_lin}), the time-reversal symmetry of the original model is preserved, 
and thus we can use the $Z_2$ topological invariants for the effective model.
The topological invariants are represented by eqs. (\ref{nu_STI}) and (\ref{nu_WTI}) in which $\epsilon_c ({\bm k})$ and $\bar{\epsilon}_f ({\bm k})$ are replaced by $\tilde{\epsilon_c} (\bm k)$ and  $\tilde{\epsilon}_f ({\bm k})$. The topological invariants include $b$ and $\lambda$, which are calculated for each temperature $T$ and laser intensity $A$. We calculate numerically $b=b(T, A)$ and $\lambda = \lambda(T, A)$ (Here we assume $A_x=A_y=A$) and use them to obtain $\nu_{\mathrm{STI}} (b(T, A), \lambda(T, A) )$ and $\nu^\alpha_{\mathrm{WTI}}(b(T, A), \lambda(T, A) )$. 
Note that we here apply the formula of the topological invariants at zero temperature to the mean-field Hamiltonian. This means that we calculate the topological invariants for the renormalized band structure determined at finite temperature. In the low temperature regime sufficiently below the Kondo temperature, which roughly corresponds to the gap size, the renormalized band structure is well defined and thus our treatment should appropriately describe the topological phases even at finite temperatures.

\begin{figure}[t]
\includegraphics[width=8.5cm]{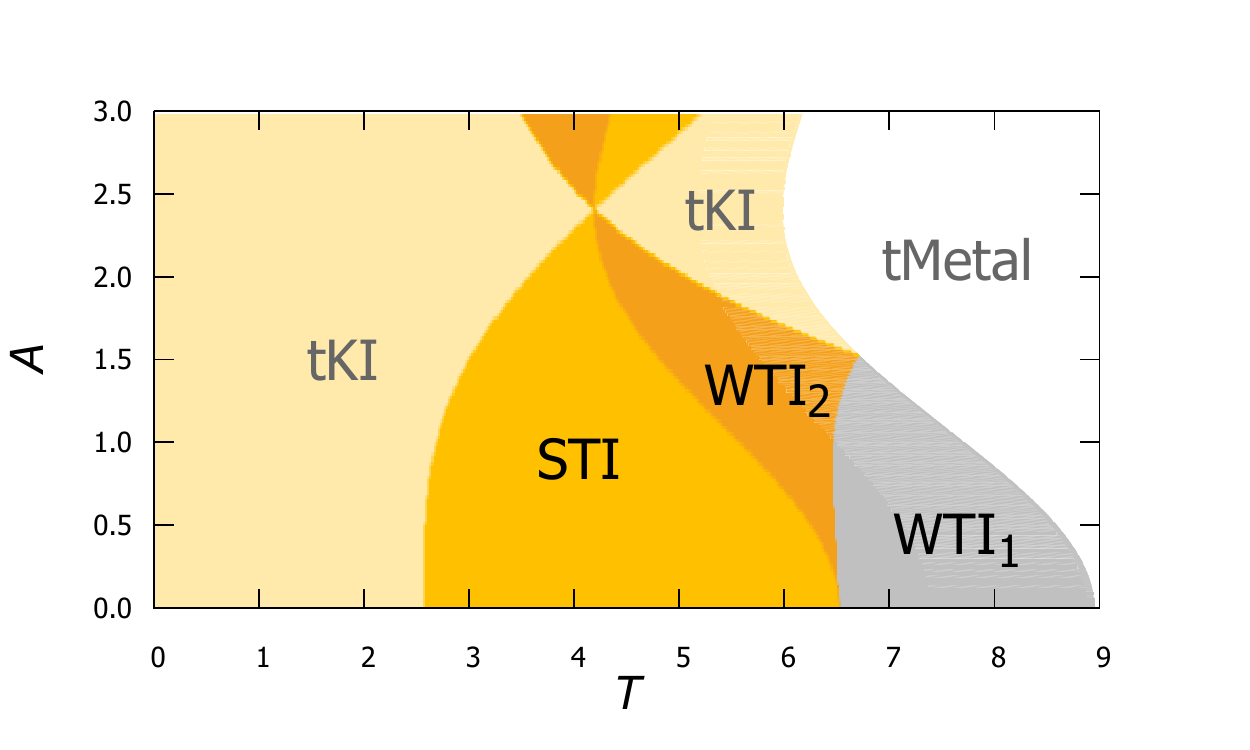}
\caption{The phase diagram of the topological Kondo insulators irradiated by linearly polarized laser light. There are two topologically trivial phases and three topologically non-trivial phases:  the trivial Kondo insulator phase (tKI, $b \neq 0$, all the topological invariants are zero), the trivial metal phase (tMetal, $b = 0$), the strong topological insulator (STI, $b\neq0$, $\nu_\mathrm{STI} = 1$) phase and two kinds of the weak topological insulator (WTI) phases, $\mathrm{WTI_1}$ ($\bm \nu = (\nu^{x}_\mathrm{WTI},\nu^{y}_\mathrm{WTI}, \nu^{z}_\mathrm{WTI})=(1,1,1)$)  and  $\mathrm{WTI_2}$ ($\bm \nu =(1,1,0)$). Here we use the parameters as $t_c=1$, $t_f=-0.2$, $V=3$, $\epsilon_f=-4$ and $\omega=12.5$.}
\label{PD_lin}
\end{figure}

First we show the numerical results of the temperature dependence of $b$ and $\lambda$ in Fig. \ref{TKI_lin} (a) and (b) respectively. 
They show that the Kondo temperature decreases with increasing the intensity of laser light. 
This behavior is similar to the off-site model discussed in Sec. IV because the model (\ref{Model_TKI}) also has only off-site hybridizations. 

Using these results, we calculate the topological invariants (\ref{nu_STI}) and (\ref{nu_WTI}) and present the results in Fig. \ref{TKI_lin_top}. 
The topological invariants indeed change with temperature and laser intensity, implying that there exist the topological phase transitions induced by linearly polarized laser light.
Next we calculate the topological invariants in a broad range of $(T, A)$ systematically and obtain the topological phase diagram shown in Fig.\ref{PD_lin}. 
Note that all the boundaries located next to the trivial metal (tMetal) phase denote the crossover induced by Kondo effect.
The other phase boundaries represent the topological phase transition which is well-defined by the renormalized band structure though it looks like a crossover in the high temperature region due to thermal fluctuations.

\begin{figure}[H]
\includegraphics[width=8.5cm]{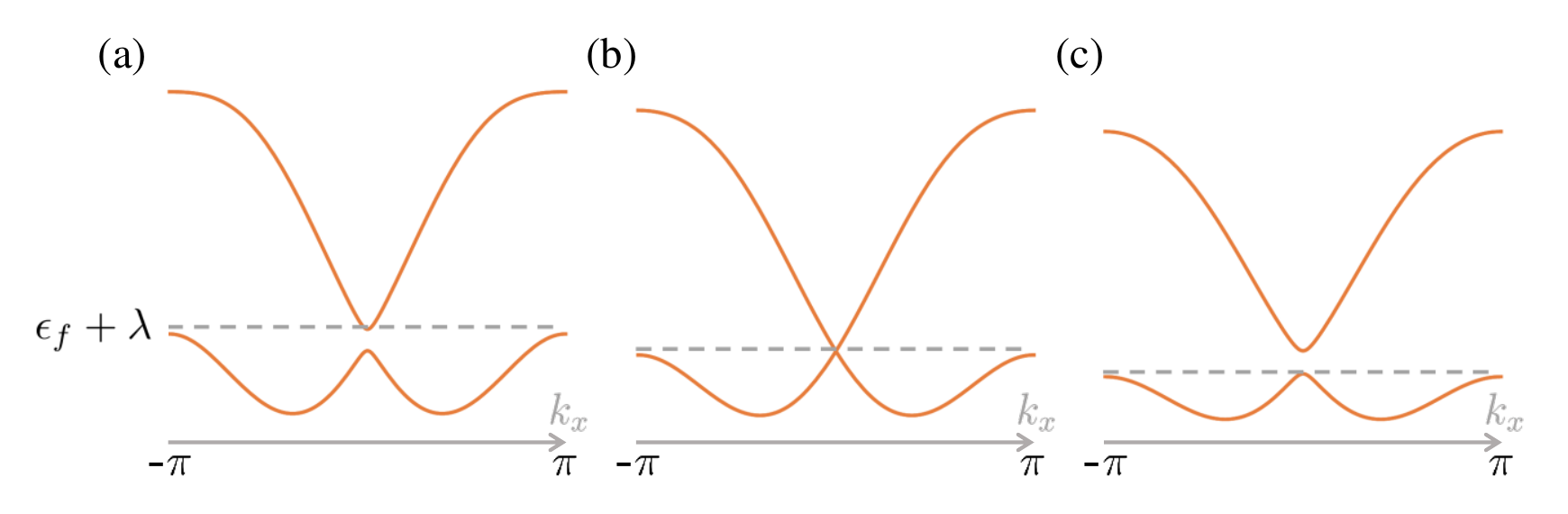}
\caption{Laser-induced topological phase transition of the topological Kondo insulators with linearly polarized laser light. We show the calculated band structure for $k_x$ (Both $k_y$ and $k_z$ are fixed to $\pi$). (a) Without laser light ($A=0.0$), the system is in the STI phase. (b) Applying the laser light ($A=0.54$), the band gap closes at the topological phase transition point. (c) Applying the strong laser light ($A=0.8$), the system is changed to $\mathrm{WTI_2}$ phase. Here we use the parameters as $T=6.2$, $t_c=1$, $t_f=-0.2$, $V=3$, $\epsilon_f=-4$ and $\omega=12.5$.}
\label{band}
\end{figure}

In Fig.\ref{PD_lin}, we find three topologically non-trivial phases: one strong topological insulator (STI) and two types of weak topological insulators (WTIs).
These WTIs have different values of the weak topological invariant of $z$-direction.
Especially the $\mathrm{WTI_2}$($\nu^{x}_\mathrm{WTI}=\nu^{y}_\mathrm{WTI}=1, \nu^{z}_\mathrm{WTI} = 0$) is the phase which appears only with a finite value of the laser intensity $A$, because the laser light breaks the cubic symmetry to the tetragonal symmetry and the appearance of $\mathrm{WTI_2}$ is permitted.
In the low intensity regime ($A \lesssim 1.0$), we find a laser-induced topological phase transition to the $\mathrm{WTI_2}$ phase near the temperature $T \sim 6.5$, which corresponds to the boundary of the topological phases without laser light.
To clarify how the phase transition occurs, we show the calculated band structure for each laser intensity in Fig. \ref{band}.
From the figure, we find that the renormalized $f$-electron level $\epsilon_f + \lambda$ shifts due to the suppression of Kondo effect by laser light.
According to this shift of energy level, the band touching occurs at the $A=0.54$ shown in Fig. \ref{band} (b). After the band touching, the energy gap opens and the topological invariant changes.

In the strong intensity region, we find the interesting behavior that the region of the topologically non-trivial phases shrinks to the point $(T, A)=(4.2,2.4)$ and after that it becomes broad again with increasing the laser intensity. 
The value of the laser intensity $A=2.4$ corresponds to the minimal zero of the zero-th Bessel function.
At that point, the hopping and hybridization in $x$- and $y$-direction vanish due to the dynamical localization and thus the system can be regarded as a one-dimensional system in $z$-direction.
Therefore the topological invariants related to $x$- and $y$-direction, i.e. $\nu_\mathrm{STI}$, $\nu^{x}_\mathrm{WTI}$, and $\nu^{y}_\mathrm{WTI}$, must be zero and then the system becomes topologically trivial.

\ \

\subsection{Circularly polarized laser light}

Next we consider the case of the circularly polarized laser ($\varphi = - \pi /2$). 
Using the results in Sec. III, we obtain the effective model as
\begin{align}
\calH_\eff&=\sum_{\bm k \sigma} \{ \tilde{\epsilon_c} (\bm k)-\mu + |b|^2 \Phi_B(\bm k) \mathrm{sgn}(\sigma) \} c^\dagger_{\bm k \sigma} c_{\bm k \sigma} \nonumber \\
&+\sum_{\bm k \sigma} \{ \tilde{\epsilon}_f ({\bm k}) + \lambda -\mu + |b|^2 \Phi_B(\bm k) \mathrm{sgn}(\sigma) \} f^\dagger_{\bm k \sigma} f_{\bm k \sigma}\nonumber \\
&+\sum_{\bm k \sigma \sigma^\prime} \{b^* ((\tilde{\bm V} (\bm k) + \bm \Phi(\bm k))\cdot \bm \sigma)_{\sigma \sigma^\prime} c^\dagger_{\bm k \sigma} f_{\bm k \sigma^\prime} + \mathrm{h.c.} \},\label{Heff_TKI_cir}
\end{align}
\begin{widetext}
with
\begin{align}
\bm \Phi(\bm k)&=-\frac{4i (t_c- |b|^2 t_f) V}{\omega}  \calJ(A_x, A_y)  (a_1 \cos k_x \sin k_y, - a_2 \sin k_x \cos k_y,0),\\
\Phi_B(\bm k) &= -\frac{2 V^2}{\omega} \calJ(A_x, A_y) (a_1 a^*_2 + a^*_1 a_2 ) \cos k_x \cos k_y. \label{laser-mag}
\end{align}\end{widetext}
$\calJ(A_x, A_y)$ has already been defined in eq. (\ref{calJ}). 
In this effective model, we obtain two terms that come from the commutator contribution $\calH_\mathrm{com}$. 
These terms break the time-reversal symmetry, as the circularly polarized laser does.
One is the laser-induced hybridization (LH) term $\bm \Phi (\bm k)$ stemming from the commutator of the kinetic term and the hybridization term. From the commutator between the $x$ and $y$ components of hybridization term, we obtain the other term $\Phi_B (\bm k)$, which plays a role of the laser-induced Zeeman coupling to (pseudo)spins.
Then we regard them as a laser-induced ``magnetic" field (LM). 
This term behaves in a similar way to a usual magnetic field, but there are two differences from the usual magnetic field.
One is that the LM depends on momenta, namely, it behaves like a Zeeman-type spin-orbit coupling. 
The other is that the LM is proportional to a square of the mean field, $|b|^2$. The mean field $|b|$ is finite only below the Kondo temperature and becomes larger in the low temperature regime. Namely, this magnetic field grows with decreasing temperature.
Especially the second difference is a quite unique feature because this behavior reflects the Kondo effect and thus it only can appear in many-body Floquet systems.

\begin{figure}[t]
\includegraphics[width=8cm]{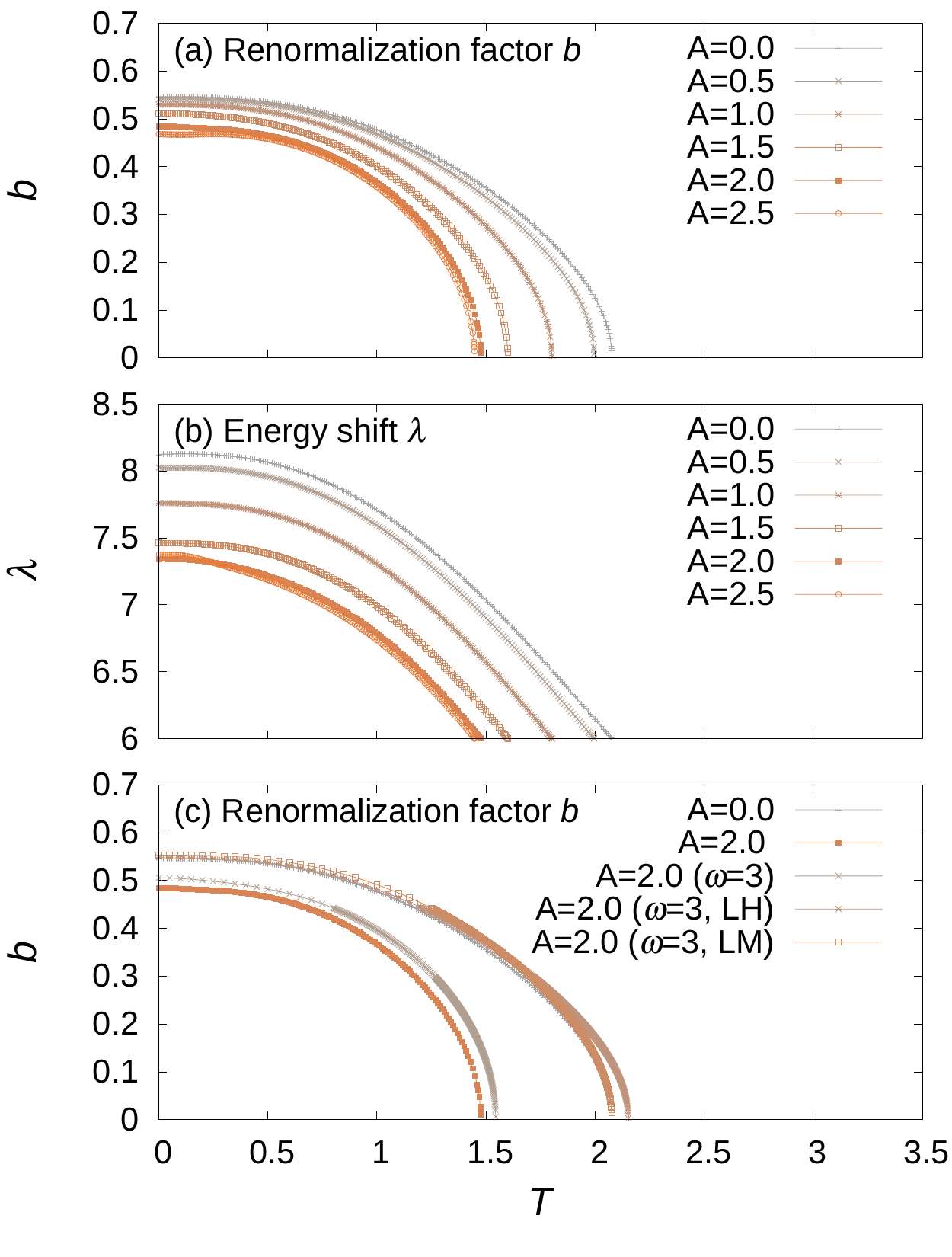}
\caption{Numerical solutions of the self-consistent equations for the model of  the topological Kondo insulator with the circularly polarized laser light. In Fig. (a) and Fig. (b), we show the temperature dependence of the renormalization factor $b$ and the energy shift $\lambda$ respectively. Here $A_x=A_y=A$ and we use the parameters as $t_c=1$, $t_f=-0.2$, $V=1.5$, $\epsilon_f=-4$ and $\omega=12.5$.  In Fig. (c), to clarify the effect which appears only in the case of circularly polarized laser, we compare the results of $\omega =3.0$ and $\omega = 12.5$ at the same intensity ($A=2.0$). LH and LM represents the case that only includes the effect of laser-induced hybridization and laser-induced magnetic field respectively. The other parameters than the frequency are the same as Fig. (a) and Fig. (b).}
\label{TKI_cir}
\end{figure}

As we mentioned above, the time-reversal symmetry is broken in this effective model (\ref{Heff_TKI_cir}).
Therefore we cannot define the original topological numbers (\ref{nu_STI}) and (\ref{nu_WTI}).
On the other hand, it is known that time-reversal symmetry or inversion symmetry broken topological insulators can be Weyl semimetals \cite{Yan2017},
which are topological gapless phases of matter recently gathering great attention because they have the exotic Fermi-arc surface states and show the exotic transport phenomena related to Chiral anormaly \cite{Hosur2013}.
Weyl semimetals are characterized by the existence of pairs of topologically non-trivial point nodes, which correspond to the monopole and anti-monopole of Berry curvature.
The texture of the Berry curvature can be detected by the Chern number for the two-dimensional subspace of the three-dimensional Brillouin zone. 

To detect the Weyl semimetallic phase, we calculate the Chern numbers for the $k_x$-$k_y$ plane (fixing $k_z$) for several values of $k_z$.
The Chern number for each $k_z$ is defined as follows:
\begin{align}
C (k_z) = \frac{1}{2 \pi i} \int_{\mathrm{B.Z.}} d k_x d k_y \epsilon^{\alpha \beta} \sum_{n \mathrm{: filled}} \partial_{k_\alpha} \braket{u_n(\bm k)| \partial_{k_\beta} u_n(\bm k)},
\end{align}
where $\alpha$ and $\beta$ run over $x$ and $y$. $\ket{u_n(\bm k)}$ is the Bloch wave function defined by the eigenfunction of the effective Hamiltonian with numerically calculated $b$ and $\lambda$ for each $(T, A)$. 
If we find the change of the Chern number $C(k_z)$ at $k^*_z$, i.e. $C(k_z > k^*_z) \neq C(k_z < k^*_z)$, we conclude that there is a Weyl node in the $k_z=k^*_z$ plane.
The change of the Chern number at $k^*_z$ corresponds to the monopole charge of the Weyl nodes.
Here we numerically calculate the Chern numbers with Fukui-Hatsugai-Suzuki method by discretizing the Brillouin zone \cite{Fukui2005}. 
Note that we apply the formula of the topological number at zero temperature to the mean-field model which can describe even the finite temperature regime. As mentioned in the previous subsection, this treatment is reasonable in the low temperature regime sufficiently below the Kondo temperature. 

Before evaluating the Chern number, we calculate $b$ and $\lambda$ for each $(T, A)$ to obtain the renormalized band structure. 
We show the numerically calculated $b$ and $\lambda$ in Fig.\ref{TKI_cir} (a) and (b). 
From the figures, we find that the Kondo temperature decreases with increasing the intensity of the laser light. 
This behavior is qualitatively same as the linearly polarized case explained in the previous subsections. 
The difference from the linearly polarized laser is only the existence of the LH and LM.
However the effect of them on the Kondo effect is very small and difficult to find out in the high frequency regime because they are the first-order correction in the $1/\omega$ expansion.
To clarify their effect, in Fig. \ref{TKI_cir} (c), we show the results in the case of $\omega = 3$, in which the $1/\omega$ expansion could not be used for quantitative arguments, but it is useful to find the qualitative effect of LH and LM. 
From Fig. \ref{TKI_cir} (c), it is seen that the LH enhances the Kondo effect in all the regime below the Kondo temperature since it enlarges the amplitude of the hybridization.
On the other hand, the effect of LM is different depending on the temperature regime. Very near the Kondo temperature, it seems that LM does not affect the original model. It is because the value of $b$ becomes small there, and thus the amplitude of LM $\sim |b|^2 V^2/\omega$ also becomes small. In the temperature below the Kondo temperature, LM slightly enhances the value of $b$, i.e. the size of the Kondo gap. This is quite different from the usual magnetic field, which is known to suppress the Kondo effect. The key difference of the LM from the usual magnetic field is the existence of the renormalization factor $|b|^2$. This factor is considered to play a important role at the low temperature and realize the enhancement of the value of $b$.  

\begin{figure}[t]
\sidesubfloat[]{\includegraphics[width=8cm]{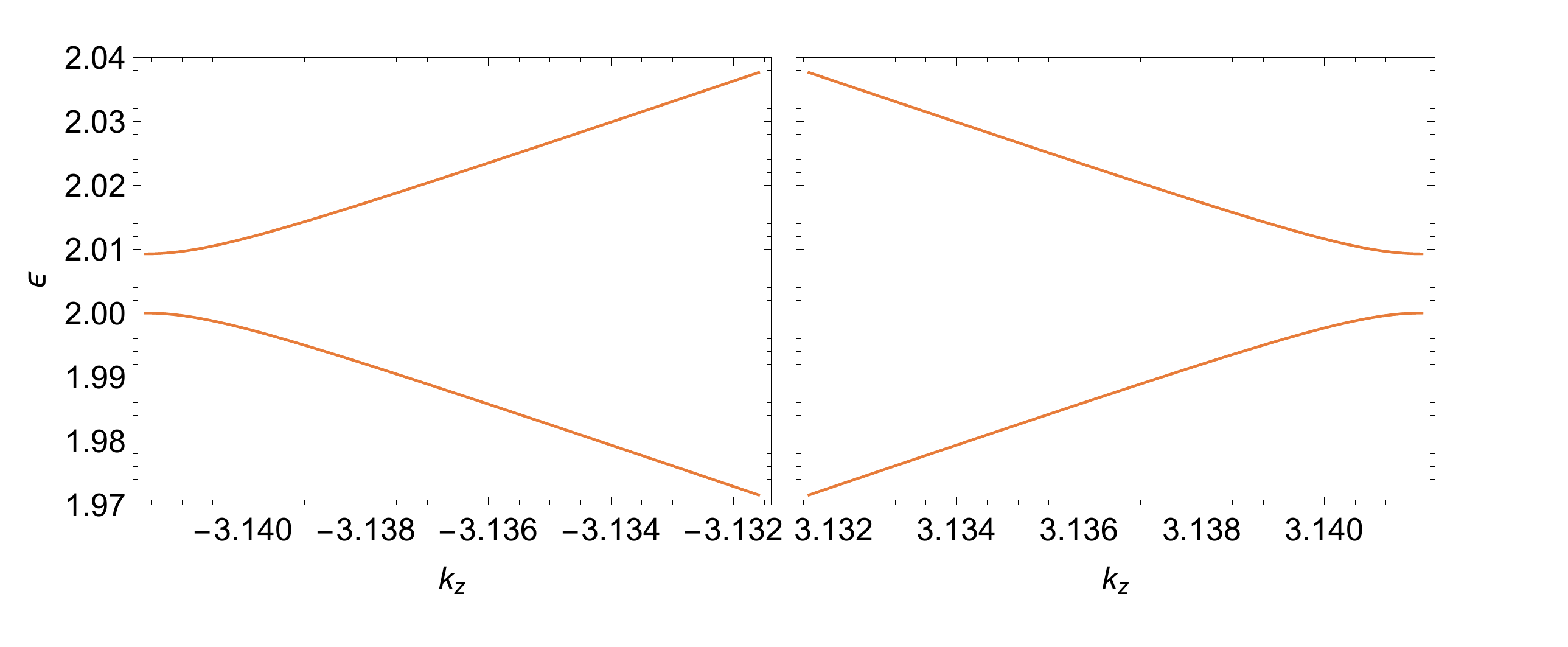} }\\
\sidesubfloat[]{\includegraphics[width=8cm]{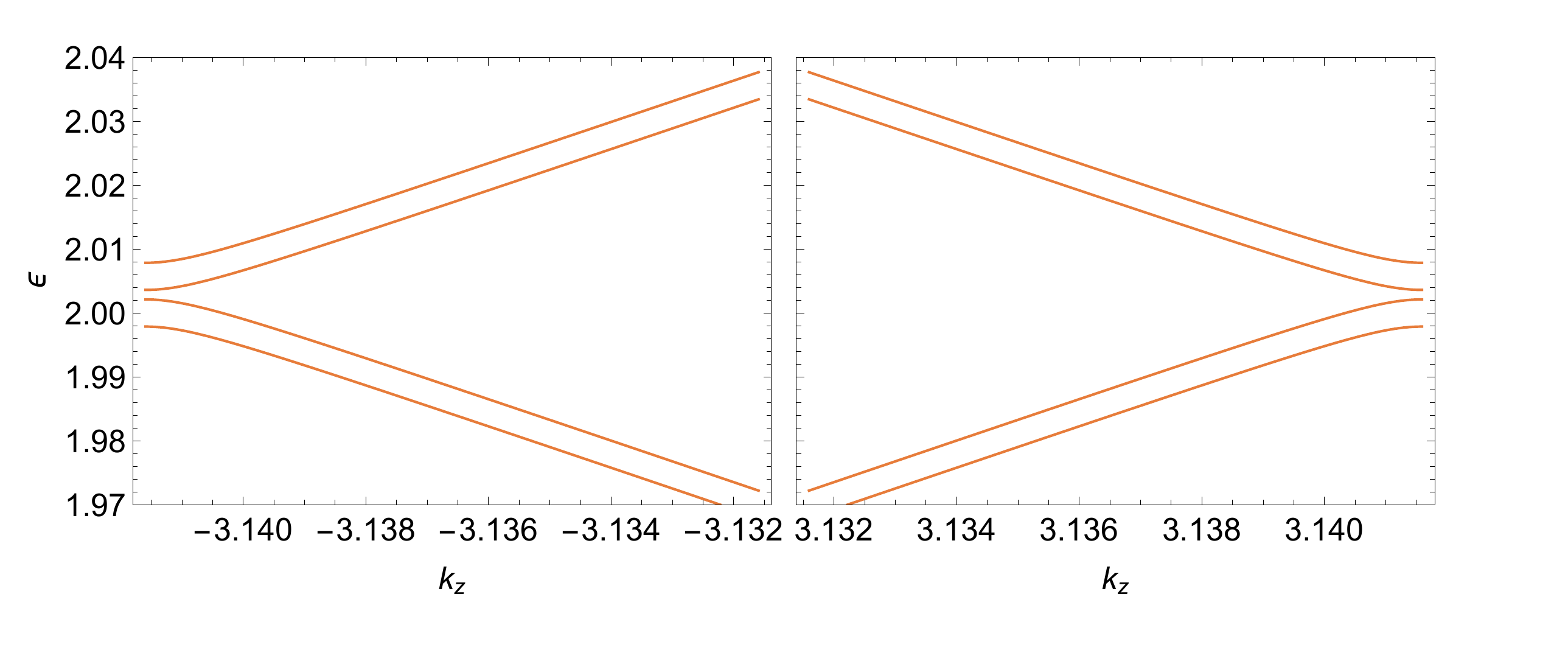} }\\
\sidesubfloat[]{\includegraphics[width=8cm]{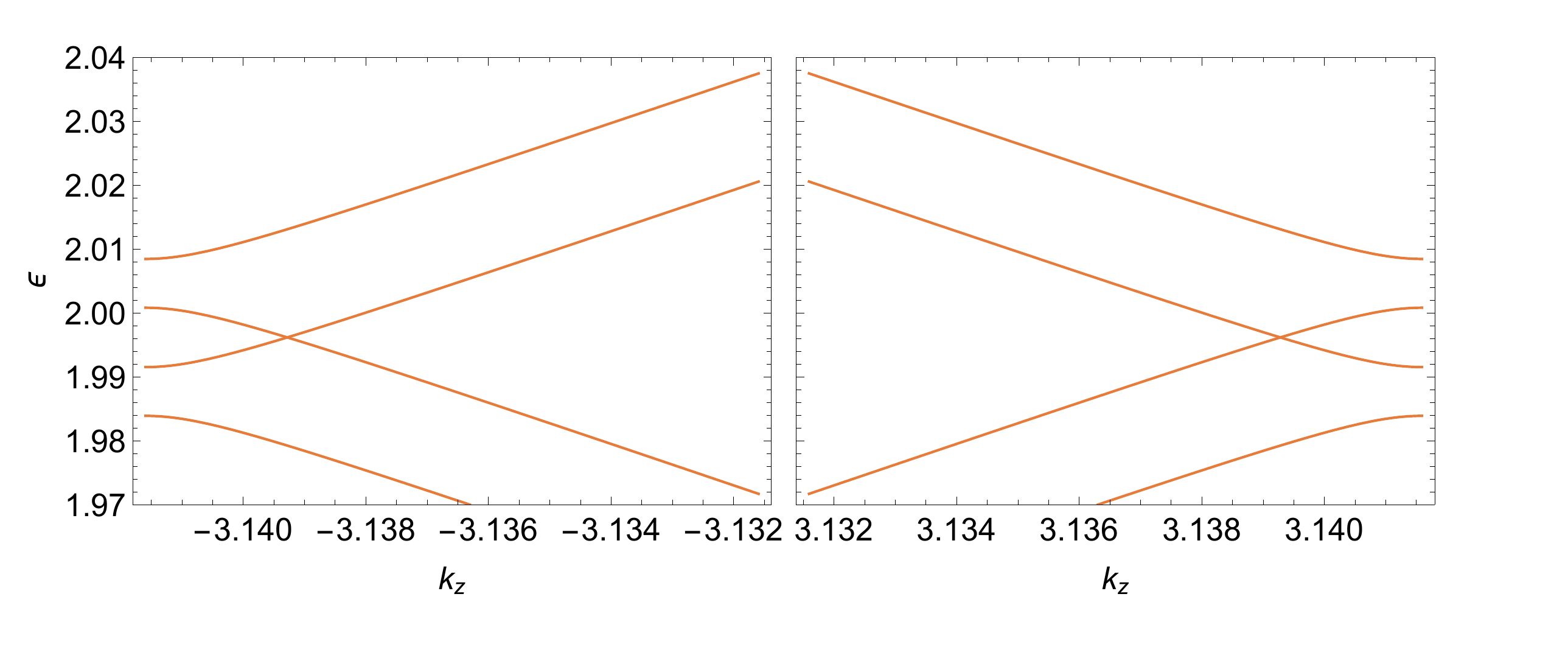} }
\caption{Laser-induced topological phase transition of the topological Kondo insulators with circularly polarized laser light.
We show the calculated band structure for $k_z$ ($(k_x, k_y)=(0,\pi)$) near the $k_z =\pm \pi$.
(a) Without laser light ($A=0.0$), the system is in the STI phase.
(b) Applying the laser light ($A=0.1$), the Kramers degeneracy is lifted and the bands are split by the LM.
(c) Making the laser light stronger (A=0.2), the crossing points (Weyl nodes) appear near $k_z = \pm  \pi$ and the system is changed to Weyl semimetallic phase
Here we use the parameters as $T=0.1$, $t_c=1$, $t_f=-0.2$, $V=1.5$, $\epsilon_f=-4$ and $\omega=12.5$.}
\label{Weyl_band}
\end{figure}

\begin{figure}[t]
\sidesubfloat[]{\includegraphics[width=8.5cm]{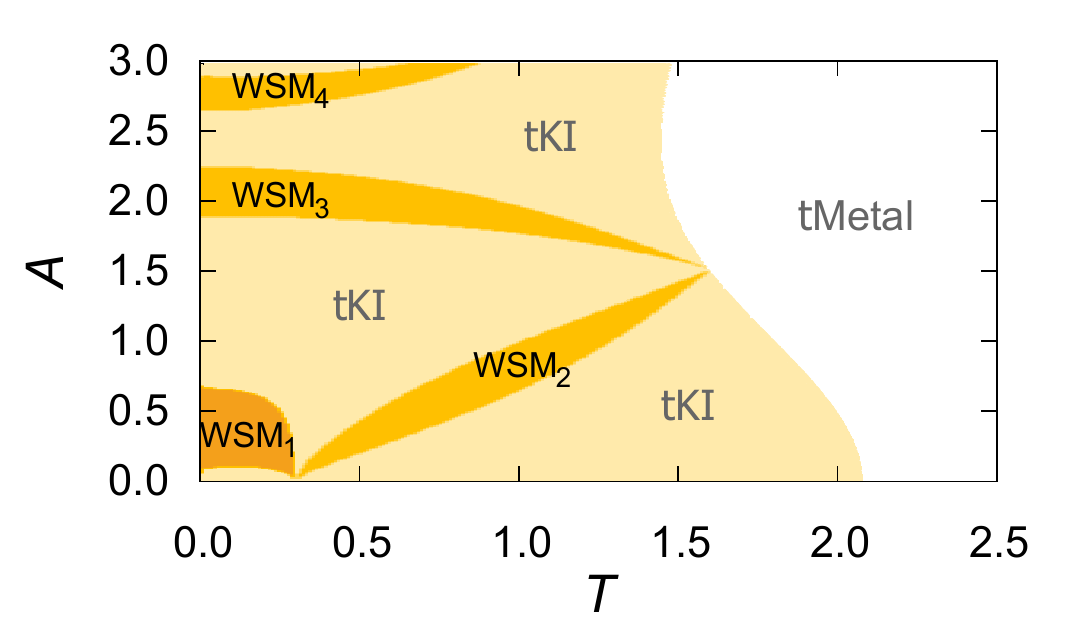}}\\
\sidesubfloat[]{\includegraphics[width=6cm]{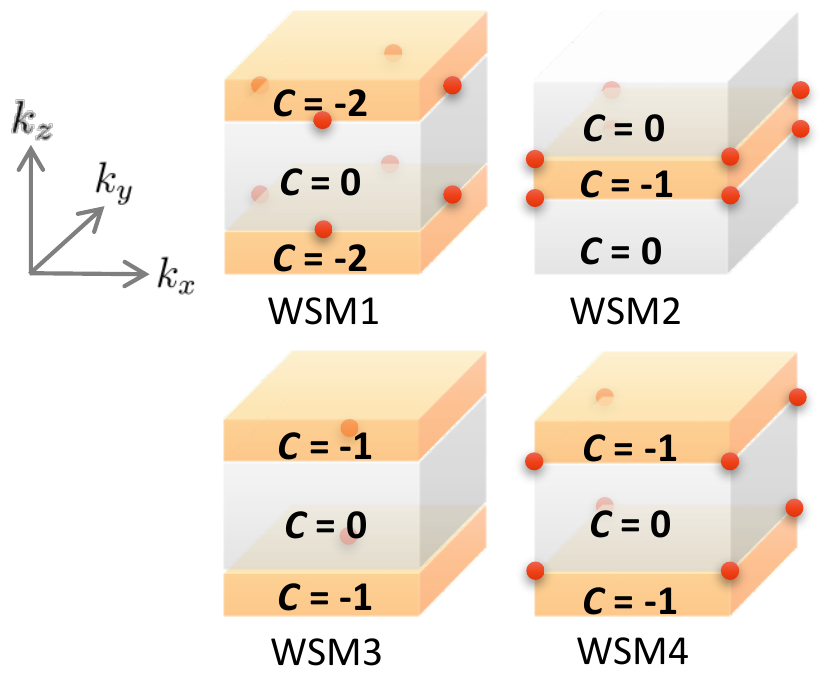}}
\caption{(a) The phase diagram of the topological Kondo insulators irradiated by circularly polarized laser light.
There are two topologically trivial phases and four topologically non-trivial phases: 
 the trivial Kondo insulator phase (tKI, $b \neq 0$, all the topological invariants are zero), the trivial metal phase (tMetal, $b = 0$), and four kinds of Weyl semimetallic phases (WSM$_1$, WSM$_2$, WSM$_3$ and WSM$_4$) shown in below. Here we use the parameters as $t_c=1$, $t_f=-0.2$, $V=1.5$, $\epsilon_f=-4$ and $\omega=12.5$. 
(b) The four kinds of Weyl semimetallic phases. We show the position of the Weyl nodes (red points) in the three-dimensional Brillouin zone. The white (orange) regions represents the Chern number $C(k_z)$ is zero (finite).}
\label{WSM_PD}
\end{figure}

Using the calculated $b$ and $\lambda$, we obtain the renormalized band structure and find the phase transition to Weyl semimetallic phases.
To clarify this point, we show the deformation of the band structure by laser light in Fig. \ref{Weyl_band}.
Before applying the laser light, we prepare the system in the STI phase ($A=0.0$, $T=0.1$).
With a finite laser intensity, the bands show the Zeeman-type splitting due to the LM.
Increasing the laser intensity, the bands touch and cross each other. The crossing point can be Weyl nodes, which characterize the Weyl semimetal. 
We calculate the change of the Chern number when we go over the crossing point in the $k_x$-$k_y$ plane, and find that the change is two, which corresponds to the two Weyl nodes at $(k_x, k_y, k_z)= (0, \pi, \pi), (\pi, 0, \pi)$ and . This evidences the topological non-triviality of the crossing points and the appearance of the Weyl semimetallic phase.

Next we calculate the Chern number for a broad range of $(T, A)$ plane and obtain the phase diagram shown in Fig. \ref{WSM_PD}(a).
We find four Weyl semimetallic phases, which have the Weyl nodes in different positions respectively.
They are summarized in Fig. \ref{WSM_PD}(b).
We can control the appearance of these Weyl semimetallic phases by changing the temperature and the intensity of the laser light.
We remark on two points about this phase diagram.
One is that Weyl semimetallic phases are realized by relatively low intensity of laser light at the temperature near $T \sim 0.3$.
The temperature $T \sim 0.3$ corresponds to the transition temperature from STI phase to $\mathrm{WTI_2}$ phase in this model.
Therefore, to find laser-induced Weyl semimetallic phase, the system near the topological phase transition is suitable.
The other is that the Weyl semimetallic phases are extended with decreasing the temperature.
It is because the LM, which is proportional to $|b|^2$, is enhanced in the low temperature region.
This implies that the low temperature is better for observing Weyl semimetallic phases.

Finally we remark on the difference of our work from the previous ones that proposed the realization of Weyl semimetallic phases by laser light.
There are several studies discussing the Weyl semimetallic phase realized by laser light using Floquet theory \cite{Wang2014, Zhang2016, Hubener2017}.
However they are limited to the free electron systems.
In our study, we consider the Kondo insulator, which is the strongly correlated insulator, and discuss the stability of the Kondo effect under the laser light.
We find that the Kondo effect is suppressed by laser light but there still exists the regime where the Kondo effect survives and the Weyl semimetallic phases can be stabilized.
The application of laser light to the topological Kondo insulators is a promising way to realize the Weyl semimetals in heavy fermion systems\cite{Yuangfeng2017}. 

\section{Experimental setups}

In the above sections, we have discussed the possibility to control the Kondo effect and the topological properties in Kondo insulators by laser light. 
In this section, we discuss the experimental setups to observe the phenomena proposed in this paper.

As for candidate materials, all the Kondo insulators are applicable to our proposal since our calculation is based on a simple model and the main results do not depend on the detail of models and approximations.
There are several materials known as Kondo insulators, e.g. SmB$_6$ and CeNiSn \cite{Riseborough2000}. They are our main targets.
Moreover we expect that our results can be applicable to most of heavy fermion systems because the periodic Anderson model is a fundamental model of heavy fermion systems. 
The key ingredient for our proposal is that the system is at low temperature and the Kondo effect occurs. In this sense, other materials in heavy fermion systems, e.g. $\mathrm{CeCu_6}$ and $\mathrm{CeCoIn_5}$ \cite{Coleman_book}, are also in the scope of our study.  Note that, as for the results about topological properties discussed in Sec. V, the system should be a topological Kondo insulator having the spin-dependent and off-site hybridization. Therefore, the candidate materials for topological Kondo insulators, e.g. SmB$_6$ \cite{Dzero2015} and YbB$_{12}$ \cite{Hagiwara2016}, are suitable to observe the laser-induced topological phase transitions.
Regarding the sample, it should be a sufficiently thin slab, because laser light only penetrates into the thin region from the surface. How deep the laser light penetrates into the sample depends on materials and laser intensity, and thus we have to prepare an adequate sample to confirm our proposal. Moreover,  the sample must be at low temperature below the Kondo temperature $T_K$, and therefore the application of laser light should be done in such low temperature regime.

The laser light is characterized by its frequency and intensity. 
Since our calculation is based on the high-frequency expansion, the frequency must be sufficiently high and off-resonant. 
Considering the energy scale in our models, frequency must be higher than the band width 12$t$. Thus the ultraviolet light is suitable for our systems. 
In addition, in solid state systems, there are many unoccupied bands above Fermi energy, then we have to choose an appropriate frequency so as to make it off-resonant to any bands. 
If the frequency is not so high or the effect of resonance is crucial, the description by the effective Hamiltonian becomes worse.
To access such a regime of frequency, we have to use different methods with which we can treat both the effect of low frequency and the interaction.
We give a remark on this point later in Sec. VII.
In regard to the intensity of the laser light, 
we need a reasonably strong intensity according to the experimental accuracy and the phenomena which we want to observe.
For example, looking at the phase diagram of the topological Kondo insulators with circularly polarized laser light in Fig. \ref{WSM_PD} (a),
we can say that the laser intensity $A$ higher than 0.1 is necessary to realized the Weyl semimetallic phase (WSM$_1$).  
The intensity $A \sim 0.1$ corresponds to the intensity of the electric field $E \sim 30$ MV/cm.

In experiments, we should use pulse laser with a finite width in time direction in order to obtain strong intensity of laser light.
In terms of the protocol of the application of laser light,
we have to consider the following two conditions. (i) Too long laser pulse drives the interacting systems to infinite-temperature trivial states \cite{DAlessio2014},
and thus we need to choose a short laser pulse. In other words, we have to stop the driving before the system completely heats up \cite{Kuwahara2016}.
 (ii) It takes time to reach the states described by the effective Hamiltonian. Moreover, to realize the ground states of the effective Hamiltonian we have to switch on the driving adiabatically \cite{Kitagawa2011, Poletti2011}. Therefore we should use a long and slowly developing pulse. 
Taking into account these two conditions, we have to choose intermediate time scale adequately. In fact, it is possible to take such an intermediate time scale. Regarding the first condition, we can make the time before heating up longer using higher frequency of laser light \cite{Mori2016}. Concerning the second condition, such a relaxation can happen in the ultrafast (femtosecond) time scale  in strongly correlated electron systems \cite{Ishikawa2014}. We should choose the pulse with sufficiently high frequency and with a longer width than the femtosecond time scale.

Finally, we address the experimental methods to confirm our proposal.
As mentioned above, we should use the pulse laser and thus have to choose a method suitable for observing transient phenomena.
Therefore transport or optical measurement is suitable. We explain how to observe the control of the Kondo effect and the topological phases respectively.
As for the Kondo effect, discussed in Sec. IV, we have shown that laser light can enhance or suppress the Kondo temperature, resulting in the shift of the temperature where the Kondo crossover occurs.
However, it is difficult to detect the shift itself since there is no singularity in the Kondo crossover.
To observe this phenomenon clearly, we can utilize the quantum phase transition in terms of the Doniach phase diagram.
As mentioned in the end of Sec. IV, the suppression or enhancement of Kondo effect changes the Kondo coupling and then the quantum phase transition occurs due to the competition between Kondo effect and RKKY interaction. The signature of this phase transition should be much more clear than the Kondo crossover.
To find the signature of the phase transition by optical measurement, the pump-probe photoemission spectroscopy (PES) is the most promising way.
We measure the system irradiated by the appropriate pump pulse, and then we can find the Kondo peak near the Fermi surface in the photoemission spectrum when the Kondo effect is dominant. On the other hand, the Kondo peak vanishes when the RKKY interaction is dominant. Therefore, performing the pump-probe PES with changing the laser intensity, we should find the appearance or vanishment of the Kondo peak when the system goes over the quantum critical point, which is the signature of the phase transition.
Concerning the topological phases, the time-resolved and angle-resolved PES is the most promising experiment tool.
If we find the change of the topological surface states, it evidences the topological phase transitions.
The closing of the bulk gap is also a signature of the topological phase transitions.  Applying linearly polarized laser light, we find the phase transitions  from STI to WTI$_2$. In this phase transition, we should find the disappearance of the surface states on the (0,0,1) surface since $\nu^z_\mathrm{WTI}$ changes from one to zero. In the case of the Weyl semimetallic phases realized by circularly polarized laser light, different Fermi-arc surface states are observed depending on which the Weyl semimetallic phases appear.

\section{Summary and Outlook}

In this paper, we have derived the effective model of Kondo insulators under high frequency laser fields with slave boson approach and Floquet theory. 
In the effective model, we have found two generic effects induced by laser light, dynamical localization and laser-induced hopping and hybridization. 
These effects change the original system and enable us to control its Kondo effect and topological properties.
Regarding Kondo effect, we have found that we can enhance or suppress the Kondo effect depending on the structure of the hybridization. 
As we discussed in the end of the Sec. IV, the enhancement and suppression of Kondo effect open a way to realize laser-induced quantum phase transitions. 
As for topological properties, we have found various topological phase transitions realizable in the topological Kondo insulators. 
With linearly polarized laser, the suppression of Kondo effect shifts the $f$-electron level and then induces the WTI phase which does not appear in the original model.
Applying the circularly polarized laser light breaks the time-reversal symmetry and gives rise to the laser-induced synthetic magnetic fields,
which creates the Weyl nodes in the band structure and realizes the Weyl semimetallic phases. 
Finally we have discussed the experimental setups to confirm our results.

We have discussed the physics in the laser-irradiated heavy fermion systems to develop the Floquet engineering in strongly correlated electron systems and
found several basic effects which change the original nature of the heavy fermion systems drastically.
However, there is room for improvement of our study from the two viewpoints.
One is the treatment of the interaction effect.
We have used the slave boson approach, which well-describes the Kondo effect itself.
However, using this approach, it is difficult to discuss the competition with magnetic or superconducting orders which are frequently observed in heavy fermion systems.
The other is the range of the frequency. Our study is based on the high-frequency expansion of the effective Hamiltonian, which is a very useful tool to clarify the qualitative behavior in the high frequency limit. Thus we cannot apply our results directly in the case of low frequency. 
To overcome these points, we have to use more sophisticated methods, with which we can treat the interaction effect and the effect of low frequency laser fields. 
For example, Floquet+DMFT formalism \cite{Tsuji2008} is known to be applicable to such a situation. In addition to these effects, we can treat the effect of bath in this method.
Applying this formalism to various systems to discuss the laser-irradiated strongly correlated systems is an important direction in the future study.

\begin{acknowledgments}
We are thankful to Takuya Nomoto and Youichi Yanase for valuable discussions. 
We also thank to Shin-ichi Kimura to give a helpful comments from the experimental point of view. This work is supported by a Grant-in-Aid for Scientific Research on Innovative Areas “Topological
Materials Science” (KAKENHI Grant No. 15H05855 ) 
and also JSPS KAKENHI (Grants No. 16J05078, 
No. 14J01328,
and No. 16K05501 
). MN is supported by RIKEN Special Postdoctoral Researcher Program.
KT and MN thank JSPS for the support from a Research Fellowship for Young Scientists.
\end{acknowledgments}

\appendix

\section{Derivation of the self-consistent equations with slave boson approach}

We show how the self-consistent equations (\ref{SCeq1_0}) and (\ref{SCeq2_0}), which determine $b$ and $\lambda$, are derived with slave boson approach.
In this approach, to treat the correlation effect in $f$-orbit, the slave particle operator $b_i$ is introduced as 
\begin{align}
f^\dagger_{i \sigma} &\rightarrow f^\dagger_{i \sigma} b_i ,\\
f_{i \sigma} &\rightarrow f_{i \sigma} b^\dagger_i,
\end{align}
with the constraint,
\begin{align}
f_{i \uparrow}^\dagger f_{i \uparrow} + f_{i \downarrow}^\dagger f_{i \downarrow} + b_i^\dagger b_i = 1. \label{const} 
\end{align}
$b_i$ ($b^\dagger_i$) is an annihilation (creation) operator of slave bosons, which corresponds to the creation (annihilation) of the holon, i.e. vacancy. The constraint (\ref{const}) represents the strong coupling limit $U \rightarrow \infty$, in which the double occupancy is completely suppressed.

With these expressions, we write down the Lagrangian corresponding to the periodic Anderson model. The constraint is taken into account with the method of Lagrange multiplier. The Lagrangian reads
\begin{widetext}
\begin{align}
\calL=
&\sum_{ij \sigma} \{ \delta_{i j} ( \partial_\tau - \mu ) + t_{c, i j} \} c^\dagger_{i \sigma}(\tau) c_{j \sigma}(\tau)+\sum_{i \sigma} \{ \partial_\tau - \mu  + \epsilon_f + i \lambda_i (\tau) \}  f^\dagger_{ i \sigma}(\tau) f_{i \sigma}(\tau) +\sum_{ij \sigma} t_{f, i j} f^\dagger_{i \sigma}(\tau)  b_{i}(\tau) b^\dagger_{ j}(\tau) f_{ j \sigma}(\tau) \nonumber \\
&+\frac{1}{\sqrt{N}} \sum_{ij \sigma \sigma'} \{ V_{i j \sigma \sigma^\prime}  c^\dagger_{i \sigma}(\tau) b^\dagger_{j}(\tau) f_{j \sigma'}(\tau) + \mathrm{h.c.} \}  + i \sum_{\bm i} \lambda_i(\tau) \{ b_i^\dagger(\tau) b_i(\tau) -1 \},
\end{align}
where $\lambda_i (\tau)$ is the Lagrange multiplier field. Next we take mean-field approximation i.e. $b_i(\tau) \rightarrow b(\tau), i \lambda_i(\tau) \rightarrow \lambda(\tau)$ and then we obtain the mean-field Lagrangian $\calL_\mathrm{MF}$ as
\begin{align}
\calL_\mathrm{MF}=
&\sum_{ij \sigma} \{ \delta_{i j} ( \partial_\tau - \mu ) + t_{c, i j} \} c^\dagger_{i \sigma}(\tau) c_{j \sigma}(\tau)
+\sum_{i \sigma}  \{ \partial_\tau - \mu  + \epsilon_f + \lambda (\tau) \} f^\dagger_{ i \sigma}(\tau) f_{i \sigma}(\tau) 
+\sum_{ij \sigma}t_{f, i j} |b(\tau)|^2 f^\dagger_{i \sigma}(\tau)  f_{ j\sigma}(\tau) \nonumber \\
&+\frac{1}{\sqrt{N}}\sum_{ij \sigma \sigma'} \{  V_{i j \sigma \sigma^\prime} b^*(\tau) c^\dagger_{i \sigma}(\tau) f_{j \sigma'}(\tau) + \mathrm{h.c.} \} 
+ \sum_{\bm i} \lambda(\tau) (|b(\tau)|^2 -1), \label{calLMF}
\end{align}
\end{widetext}
To derive the self-consistent equations for $b$ and $\lambda$, we integrate out the fermionic degrees of freedom $c$ and $f$ and derive the effective action for bosonic fields $b(\tau)$ and scalar fields $\lambda(\tau)$. The effective action $S_\eff$ is defined as
\begin{align}
e^{-S_\eff [ b^\dagger,b,\lambda]}\nonumber \equiv \int \calD [ c^\dagger, c, f^\dagger,f] e^{-S_{\mathrm{MF}} [ c^\dagger, c, f^\dagger,f,b^\dagger,b,\lambda] },
\end{align}
where
\begin{align}
S_{\mathrm{MF}} = \int_0^\beta d \tau \calL_\mathrm{MF}[ c^\dagger(\tau), c(\tau), f^\dagger(\tau),f(\tau),b^\dagger(\tau),b(\tau),\lambda(\tau)].
\end{align}
The self-consistent equations are given by the saddle point conditions for the effective action:
\begin{align}
\frac{\delta S_\eff}{\delta b(\tau)} = 0, \frac{\delta S_\eff}{\delta \lambda(\tau)} = 0.
\end{align}
These conditions can be rewritten using $S_\mathrm{MF}$ as
\begin{align}
\left \langle \frac{\delta S_\mathrm{MF}}{\delta b(\tau)} \right \rangle = 0, \left \langle \frac{\delta S_\mathrm{MF}}{\delta \lambda(\tau)} \right \rangle = 0, \label{saddle2}
\end{align}
where $\langle \cdots \rangle $ is the thermal average defined as
\begin{align}
\langle \cdots \rangle =\frac{\displaystyle  \int \calD [ c^\dagger(\tau), c(\tau), f^\dagger(\tau),f(\tau)] \cdots e^{-S_{\mathrm{MF}}}}{\displaystyle \int \calD [ c^\dagger(\tau), c(\tau), f^\dagger(\tau),f(\tau)] e^{-S_{\mathrm{MF}}}}.
\end{align}
Using eqs. (\ref{calLMF}) and (\ref{saddle2}), we can derive the self-consistent equations (\ref{SCeq1_0}) and (\ref{SCeq2_0}).

\bibliographystyle{apsrev4-1}
\bibliography{ref.bib}

\end{document}